\documentclass[runningheads]{svmult}
\usepackage{makeidx}   
\usepackage{graphicx}  
\usepackage{subeqnar}  
\usepackage{multicol}  
\usepackage{cropmark} 
\usepackage{physprbb}  
\def\H{H\hskip-8.5pt/\hskip2pt}

\def\coeff#1#2{{\textstyle{#1\over #2}}}

\def\VEV#1{\left\langle #1\right\rangle}

\def\lsim{\mathrel{\mathpalette\@versim<}}

\def\gsim{\mathrel{\mathpalette\@versim>}}

\def\Tr{{\rm Tr}\,}

\begin{document}
\title*{On CPT Symmetry: Cosmological, Quantum-Gravitational 
and other possible violations and their phenomenology}
\toctitle{On CPT Symmetry: Cosmological, Quantum-Gravitational, 
\protect\newline and other possible 
violations and their phenomenology}
%
%
\titlerunning{CPT Violation}
%
\author{Nick E. Mavromatos}
%
\authorrunning{Nick E. Mavromatos}
%
%
\institute{King's College London, Department of 
Physics-Theoretical Physics, \\
Strand, London WC2R 2LS, U.K, and \\
Departamento de F\'isica T\'eorica, Universidad de Valencia,\\
E-46100, Burjassot, Valencia, Spain.} 

\maketitle              

\begin{abstract}
I discuss various ways in which CPT symmetry may be violated, and 
their phenomenology in current or immediate future 
experimental facilities, both terrestrial and astrophysical. 
Specifically, I discuss first violations of CPT symmetry due to
the impossibility of defining
a scattering matrix as a 
consequence of the 
existence of microscopic or macroscopic
space-time boundaries, such as Planck-scale Black-Hole (event) 
horizons, or 
cosmological horizons due to the presence of a (positive) cosmological 
constant in the Universe. 
Second, I discuss CPT violation due to breaking of Lorentz
symmetry, which may characterize certain approaches to 
quantum gravity, 
and third, I describe models 
of CPT non invariance due to violations of locality of interactions.
In each of the above categories I discuss experimental sensitivities.
I argue that the majority of Lorentz-violating cases of CPT breaking, 
with minimal (linear) 
suppression by the Planck-mass scale, are already excluded 
by current experimental tests. There are however some 
(stringy) models which can evade these constraints. 
\end{abstract}

\section{CPT Breaking and the Scattering Matrix}

\begin{figure}[t]
\begin{center}
\includegraphics[width=.6\textwidth]{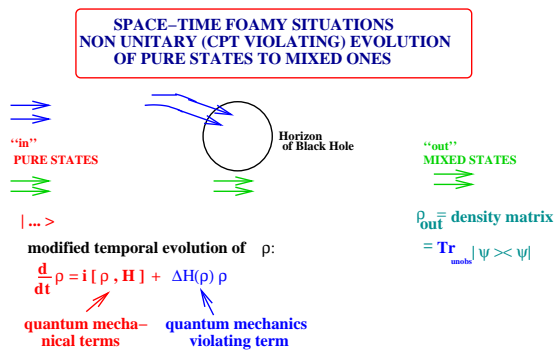}
\end{center}
\caption[]{When matter states pass by an evaporating 
Black Hole, information is lost inside the horizon. In a quantum 
gravity context, such black holes may appear as fluctuations 
of the geometry of microscopic (Planck) size, which results in 
the evolution of pure states to mixed for asymptotic observers, 
and hence in unitarity loss, and consequently CPT violation.}
\label{bhfoam}
\end{figure}

The symmetry under the successive operations (in any order) 
of charge conjugation, C, parity (reflexion), P, and 
time reversal, T, known as CPT, is a fundamental symmetry of any
\emph{local} quantum field theory in \emph{flat} space time, under the 
following assumptions~\cite{cpt}:
\begin{itemize}
\item{(i)} Unitarity and the proper definition of a scattering matrix, 
\item{(ii)} Lorentz Invariance and
\item{(iii)} Locality of Interactions
\end{itemize}
In the presence of gravity, i.e. non-Minkowski,non-flat space time 
backgrounds, CPT symmetry may be \emph{violated}, at least in 
\emph{in its strong form}. This is indeed the case of \emph{singular} 
space-time gravitational backgrounds, such as black holes, or 
in general space times with boundaries. The reason is 
that in such cases the presence of these boundaries jeopardizes 
requirements (i) and (ii) of the CPT theorem. In a quantum context, 
a Black hole evaporates due to Hawking radiation, and as such it may `capture'
for ever information on matter states passing nearby, as depicted 
schematically in 
figure \ref{bhfoam}.

In such a case, one may not be able to define properly asymptotic 
state vectors $|\psi>$ in a quantum context, given that Planck size
black hole horizons appear as fluctuations of the geometry, and hence 
an asymptotic observer will necessarily trace 
out the information captured by the horizons. This means that the out 
states will be necessarily described by 
\emph{density matrices}, $\rho = {\rm Tr}_{\rm unobs}|\psi><\psi|$.
One has therefore an  evolution from pure states to mixed, 
and unitarity is lost. The problem in defining asymptotic
state vectors also implies the impossibility of defining a proper 
scattering $S$-matrix, since the latter connects 
by definition ``in'' and ``out'' state 
vectors: $ |OUT> = S|IN>$. Instead, one 
can only define a Hawking \$-matrix~\cite{dollar}, 
which connects
IN and OUT mixed states described by \emph{density matrices}:
\begin{eqnarray} 
\rho_{OUT} = \$ \rho_{IN} 
\label{dollarmatrix} 
\end{eqnarray}
where \$ $\ne SS^\dagger$, with $S=e^{iHt}$ the S-matrix, and 
$H$ the Hamiltonian of the matter subsystem. The \$-matrix \emph{has no inverse},
as a consequence of the loss of information 
encountered in the problem.

This, in turn, results~\cite{wald} in the impossibility of defining 
a proper CPT operator for such a system, and hence CPT symmetry is violated in its strong form. The proof is elementary but instructive, and hence we give it 
here for completeness. Consider an initial ($t \to -\infty$) 
density matrix, $\rho_{IN}$, which may or may not correspond to 
a pure state. Let $\rho_{OUT}$ be the corresponding OUT state 
($t \to +\infty$), 
which is definitely a mixed state in the case at hand 
(c.f. figure \ref{bhfoam}). Assume that there exists a unitary, 
invertible quantum-mechanical CPT operator $\Theta$ : 
\begin{eqnarray} 
&& \Theta\rho_{IN} = {\overline \rho}_{OUT}~, \qquad 
\rho_{OUT} = \Theta^{-1} {\overline \rho}_{IN} \nonumber \\
&& \$ {\overline \rho}_{IN} = {\overline \rho}_{OUT}
\label{thetaope}
\end{eqnarray} 
From (\ref{dollarmatrix}), (\ref{thetaope}) 
we may write: $\$ \Theta^{-1} \rho_{OUT} = \Theta \rho_{IN} $, from which:
$ \Theta^{-1} \$ \Theta^{-1} \rho_{OUT} =\rho_{IN} $.However, from 
(\ref{dollarmatrix}) we can rewrite this equation in the form:
\begin{eqnarray} 
\Theta^{-1} \$ \Theta^{-1} \$ \rho_{IN} = \rho_{IN} 
\label{dollarinvertible} 
\end{eqnarray} 
which implies the existence of an inverse $\Theta^{-1} \$ \Theta^{-1}$
of the \$-matrix, which as explained above does not exist 
as a result of the information loss in the problem.
Thus, we conclude that in the black hole case of figure \ref{bhfoam},
or any other case of an open quantum mechanical system with 
unitarity (information) loss, a strong form of 
CPT invariance \emph{cannot hold}~\cite{wald}.

\section{Cosmological Constant, \$-matrix and CPT Breaking} 

The conclusions of the previous section 
may be extended to cases of interest in astrophysics, involving 
space-time boundaries across which (quantum) information 
is lost. Such a case is that of a Robertson-Walker expanding Universe 
with a positive cosmological \emph{constant} $\Lambda >0$. 
In such a Universe the density 
of matter scales with the scale factor $a(t)$ as: $\rho \sim a^{-3}$,
while the presence of a cosmological constant results in a 
``vacuum energy density'' $\rho_\Lambda$ that remains \emph{constant}, and hence eventually 
\emph{dominates} the expansion of the Universe, forcing the latter to 
enter a (de Sitter) phase of exponential expansion, $a(t \to \infty) \sim e^{\sqrt{\frac{\Lambda}{3}}t}$, and hence \emph{eternal} acceleration. 
In such a case there is a \emph{cosmic} horizon, i.e. a surface beyond which 
a cosmological observer in our Universe cannot see, given that the light takes 
infinite time to traverse the (finite) distance 
corresponding to the horizon radius $\delta$ :
\begin{eqnarray} 
\delta = \int_{t_0}^{\infty} = \frac{c dt}{a(t)} < \infty 
\label{horizon} 
\end{eqnarray} 
The presence of a cosmic horizon makes the case of an eternally accelerating 
Universe with a (positive) cosmological constant $\Lambda$ somewhat similar to 
the Black Hole case of figure \ref{bhfoam}. The important difference
is that in the $\Lambda$-Universe the observer lives \emph{inside} the 
cosmic (Hubble) horizon,
in contrast to the Black Hole case, where the observer lives\emph{outside}
the event horizon. Nevertheless, in \emph{both} cases, one cannot define 
proper asymptotic states, 
in the sense of information loss across 
the space-time boundary , and hence the \$ matrix is non factorisable: an 
S-matrix cannot be defined for the problem.According to the discussion 
in the previous section, therefore, one expects~\cite{wald} a 
breaking of CPT symmetry in its strong form. 
IN this case, we call this type of breaking  \emph{Cosmological},
to distinguish its global nature from the microscopic black hole case
of figure \ref{bhfoam}, which we refer to as \emph{local quantum gravity} 
type of CPT breaking.

\begin{figure}[t]
\begin{center}
\includegraphics[width=.3\textwidth]{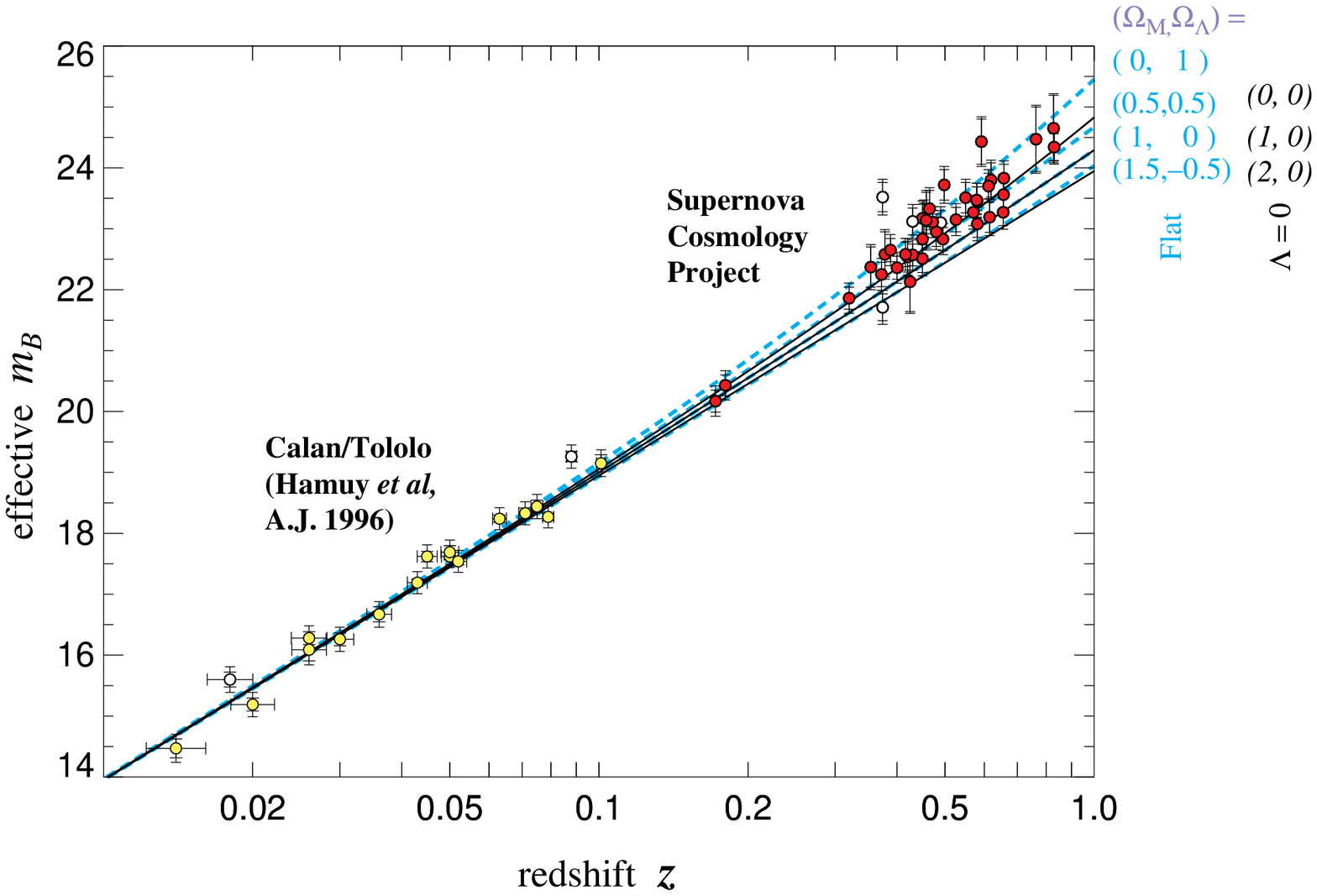}
\hfill \includegraphics[width=.3\textwidth]{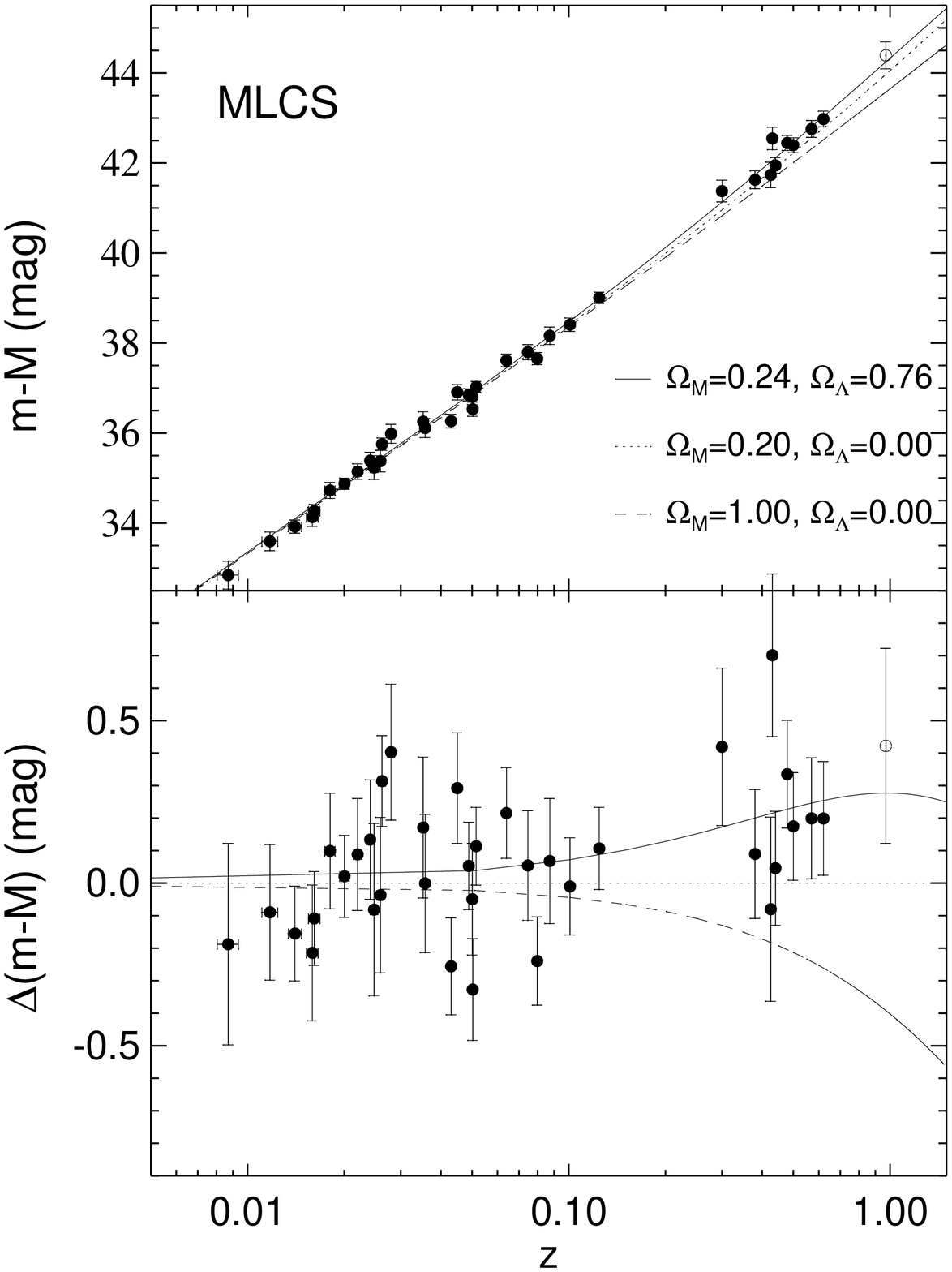}
\hfill \includegraphics[angle=90,width=.3\textwidth]{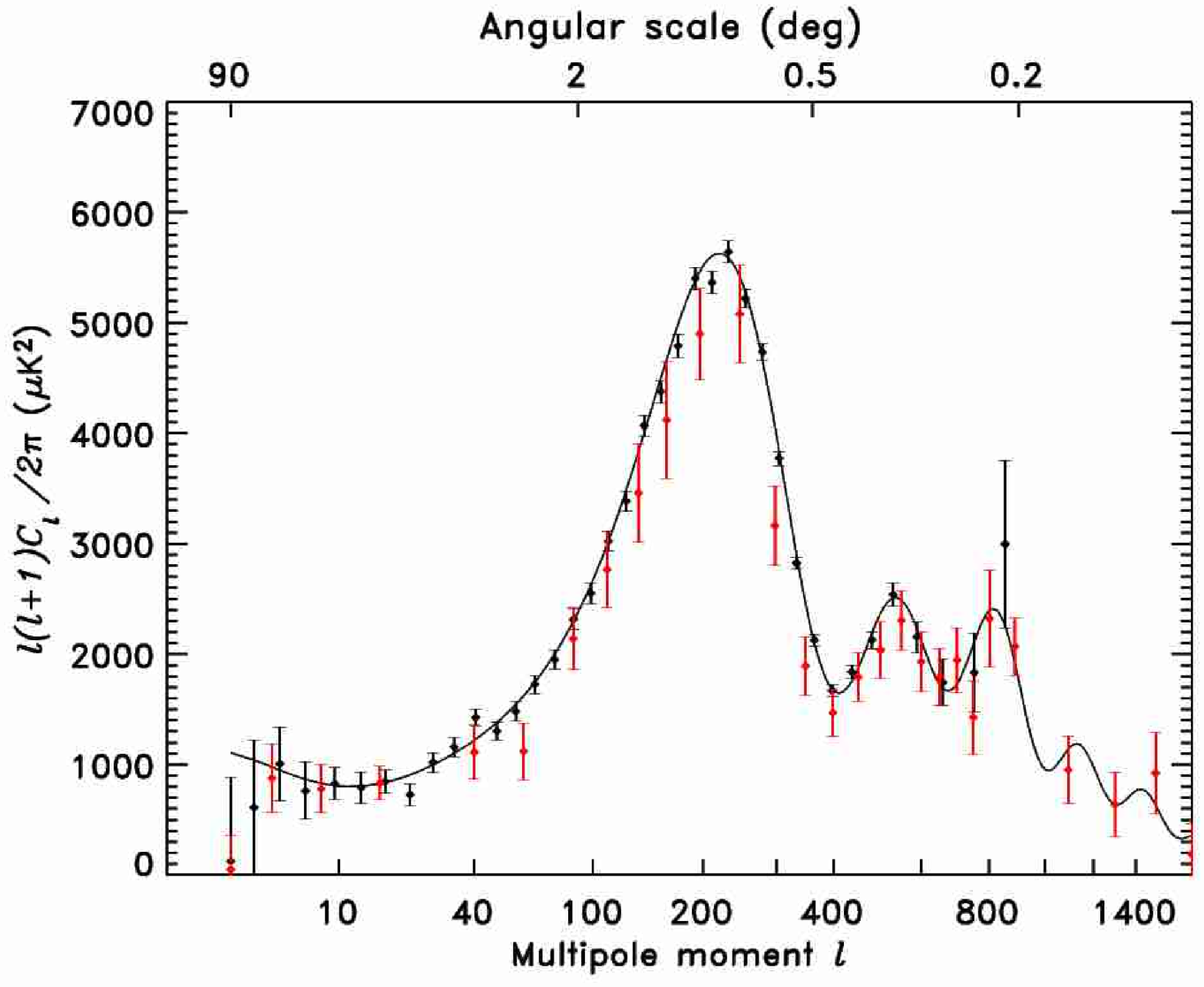}
\end{center}
\caption[]{\underline{Left and Middle Figures}: Supernovae Ia data
on the current acceleration of the Universe.
\underline{Right Figure}: Cosmic Microwave Background 
Data also point towards a current acceleration in the sense of a non-zero
vacuum energy density of the Universe, covering 73\% of its energy content.}
\label{sncmb}
\end{figure}

It must be mentioned at this stage that current astrophysical 
evidence (either by direct measurements of the acceleration of the Universe using 
supernovae type -Ia data~\cite{snIa} (c.f. left and middle figures 
\ref{sncmb}) or 
indirect evidence from Cosmic Microwave Background  (CMB) anisotropies~\cite{cmb}
(c.f. right figure \ref{sncmb}) ) points towards a \emph{dark} energy 
component of the Universe, which covers 73\%
of its energy content. Best-fit models to the current data 
are consistent with a Cosmological Constant $\Lambda$-Universe~\cite{snIa,wmap}.

It is physically interesting, therefore, to ask what happens with 
CPT invariance of a quantum field theory when placed in such 
Universes (global gravitational backgrounds). 
From the previous arguments~\cite{wald} one expects 
Cosmological Breaking of the CPT theorem. An immediate question 
concerns the \emph{order of magnitude} of such a breaking. Certainly it is expected to be 
proportional to the cosmological constant $\Lambda$, but since the latter is a
dimensionful quantity the order of the associated CPT violation 
should depend on the specific context considered. The natural framework 
will be to consider the time evolution of the density matrix $\rho_m$ of 
matter in a $\Lambda$-Universe. The analogy 
with open system quantum mechanics, as well as the Black Hole case (c.f. fig.\ref{bhfoam}), implies the following modified evolution equation:
\begin{eqnarray} 
\partial_t \rho_m = i[\rho_m, H ]+\delta H(\rho_m)
\label{evolution} 
\end{eqnarray} 
where $H$ is the Hamiltonian of the matter subsystem, 
and $\delta H(\rho_m)$ is in general non linear, and expresses 
the interaction of the matter subsystem with the ``environment''.
In the cosmological case this environment consists of long-wavelength 
modes (or parts thereof) which lie outside the Hubble cosmic horizon (\ref{horizon}). To leading order one may assume approximately that terms linear 
in $\rho_m$ in $\delta H(\rho_m)$ determine to a satisfactory degree 
the low-energy physics, accessible to particle physics experiments. Then, 
linear evolutions of the type (\ref{evolution}) may be restricted
according to the Lindblad form~\cite{lindblad}:
\begin{eqnarray} 
\partial_t \rho_m = i [\rho_m, H] + 
\sum_{j}a_j \rho_m a_j^\dagger - \frac{1}{2}\left(a_j^\dagger a_j \rho 
+ \rho a_j^\dagger a_j \right)
\label{lindblad}
\end{eqnarray}
with $a_j$,$a_j^\dagger$ operators expressing 
the interaction with the environment (the set may even be infinite). 
The form (\ref{lindblad}) 
is valid  
upon some reasonable 
assumptions about positivity of $\rho_m$, conservation of its trace
${\rm Tr}\rho_m $=const, \emph{etc.} Such assumptions may not be valid 
in some models of quantum gravity, but one may be confident that 
they at least  
characterize the cosmological constant Universe under consideration. 
To adapt (\ref{lindblad}) in a cosmological context one needs a specific
and detailed model. 

\section{Non-Critical Strings, Cosmological  Constant and the order of 
CPT Violation}

One such model, to which we shall restrict our attention
in this review, is provided within the framework of the so-called
non-critical (Liouville) string theory~\cite{ddk}, which - as argued in 
\cite{emn} - is an appropriate framework for describing 
\emph{non-equilibrium} string theories. The immediate question, of course, 
is whether the $\Lambda$-Universe is a non equilibrium system.

The answer to such a question depends on the definition of ``equilibrium''
in string theory, which should  not be confused with the `thermal' one.
In the first 
quantized version of 
strings we define~\cite{emn} 
as `equilibrium theories' the stringy $\sigma$-models 
whose target-space backgrounds correspond to 
\emph{conformal} world-sheet couplings, in other words 
have vanishing world-sheet renormalization-group (RG) 
$\beta$-functions~\footnote{Actually, diffeomorphism invariance of the target space requires~\cite{tseytlin} 
that the $\beta^i$ in (\ref{weylcoeff}) are the so-called
Weyl-anomaly coefficients and not simply the RG $\beta$-functions 
$\beta_{RG}=dg^i/d{\rm ln}\mu$, with ${\rm ln}\mu$ a world-sheet scale.
One has: $\beta^i = 
\beta_{RG}^i + \delta g^i$ 
where  
$\delta g^i$ correspond to variations of the background fields $g^i$ 
under target-space diffeomorphisms. For instance, for graviton 
backgrounds one has to ${\cal O}(\alpha ')$, with $\alpha '$ the Regge slope: $\beta_{\mu\nu}^G = \alpha ' R_{\mu\nu} + \nabla_{(\mu} \partial_{\nu)}
W[g]$, where $W[g]$ is a scalar function of the backgrounds. To 
${\cal O}(\alpha')$, $W[g]$ is the dilaton $\Phi$.}:
\begin{eqnarray} 
\beta^i = \sum_{\{i_m\}}C^{i}_{i_1 \dots i_m}g^{i_1}\dots g^{i_m} = 0 
\label{weylcoeff}
\end{eqnarray} 
where $C_{i_1 \dots i_m}$ correspond to correlation functions 
of vertex operators of the $\sigma$-model, and $g^i$ are 
$\sigma$-model couplings/target-space background fields. 
The theory space indices ${i_j}$ are raised and lowered by a 
specific metric to be discussed below.  
Such equations 
determine specific (conformal) string backgrounds. As target space equations,
they can be expanded in configuration space in powers of 
$\alpha '=\ell_s^2=1/M_s^2$, where $M_s$ the string mass scale. This 
is a free parameter in string theory, and, thus, may be different 
from the four-dimensional Planck mass $M_P \sim 10^{19}$ GeV.
 
To lowest order in $\alpha'$, in the case of a gravitational background only, 
the respective $\beta$-function is just the Ricci tensor:
\begin{eqnarray} 
\beta_{\mu\nu}^G = \alpha ' R_{\mu\nu} 
\label{ricci}
\end{eqnarray} 
where $\mu,\nu$ are space time indices.Thus, if other backgrounds are ignored,
the condition (\ref{weylcoeff}) implies Ricci flat backgrounds.
Such backgrounds are solutions of Einstein equations, which thus reconciles
 General Relativity with stringy Conformal Invariance, at least at the 
level of equations of motion in target space~\cite{tseytlin}.
The equivalence is more general, given that for arbitrary backgrounds $\{ g^i \}$
of a stringy $\sigma$-model one can prove~\cite{osborn} 
a \emph{gradient flow} property of the $\beta^i$-functions:
\begin{eqnarray} 
\beta^i = {\cal G}^{ij}\frac{\delta S[g]}{\delta g^j}~, \qquad 
{\cal G}^{ij} \equiv z^2 {\bar z}^2 <V^i V^j>_g
\label{gradientflow}
\end{eqnarray} 
where $z, {\bar z}$ are (Euclidean) world-sheet coordinates, 
${\cal G}^{ij}$ is the Zamolodchikov (inverse) metric in theory space
$\{ g^i \}$, and $S[g]$ is a target-space 
diffeomorphism invariant functional of $g$, playing the r\^ole of 
an effective action. Since, at least perturbatively
in (weak) couplings $g^i$, the function ${\cal G}^{ij}$ is invertible,  
one concludes that the conformal invariance conditions 
(\ref{weylcoeff}) are, at least perturbatively in $g^i$, 
\emph{equivalent} to on-shell equations of motion 
$\frac{\delta S[g]}{\delta g^i} =0$. When such on-shell conditions are 
satisfied we speak of `equilibrium points' in string theory space. 

The issue of \emph{non-critical} strings arises in connexion 
with non-vanishing $\beta^i \ne 0$, which in view of the above considerations
also implies off shell backgrounds in the sense of 
$\frac{\delta S[g]}{\delta g^i} \ne 0$.
This is what we consider~\cite{emn} as a \emph{non-equilibrium} case in string 
theory.

In the latter case, conformal invariance of the $\sigma$-model 
may be \emph{restored} by the introduction of an \emph{extra}
$\sigma$-model coordinate, the Liouville field~\cite{ddk}, 
$\varphi (z,{\bar z})$, the zero-mode of which may be viewed~\cite{emn} 
as the world-sheet RG scale.The important point is that upon 
Liouville dressing of the non-conformal vertex operators~\cite{ddk}, 
the stringy $\sigma$-model in the extended ((D+1)-dimensional) 
target space $(X^\mu, \varphi)$, $\mu=0,\dots, D-1,$ is \emph{conformal}. 

A detailed analysis~\cite{ddk}, 
shows that near a fixed point in theory space $\{ g^[ \}$
(where $\sigma$-model perturbation theory is valid) 
the Liouville dressed~\cite{ddk} couplings $\lambda^i(\varphi, g^j)$ 
`obey the following equation:
\begin{eqnarray} 
{\lambda^i}'' + Q[\lambda^j,\varphi] {\lambda^i}' = -\beta^i(\lambda^j)
\label{liouvcond}
\end{eqnarray} 
where the prime denotes differentiation with respect to the world-sheet
zero mode of $\varphi$, 
$\beta^i(\lambda^j)$ is the world-sheet Weyl anomaly coefficient
but with $g^i$ replaced by the Liouville dressed couplings $\lambda^j$,
and the minus sign in front of it is valid for \emph{supercritical}
string theories, i.e. for deformed $\sigma$-models with the running central
charge $C[g] > 9$ (for superstrings), where we shall concentrate on
for the purposes of this review. In this case the Liouville mode has a 
\emph{time-like} signature~\cite{aben}, and we interpret it as target time~\cite{emn}.

A solution to order $g^2$ of (\ref{liouvcond}) reads: 
\begin{eqnarray} 
\lambda^i(g^j,\varphi) = e^{-\alpha^i\varphi}g^i 
+ C^i_{jk}g^jg^ke^{-\alpha^i\varphi}\varphi + {\cal O}(g^3)
\label{liouvcondsol}
\end{eqnarray} 
with $C^i_{jk}$ the O.P.E. appearing in the $\beta$-functions
(c.f. (\ref{weylcoeff})).  The quantities $\alpha^i$ 
are the gravitational anomalous dimensions~\cite{ddk}, which are defined as follows: if we expand $Q^2[\lambda] = Q_*^2 + {\cal O}(g^2)$, where $Q_*^2 =$const,
then $\alpha^i$ satisfy: $\alpha^i(\alpha^i + Q_*)=-(\Delta_i - 2)$,
with $\Delta_i$ the conformal dimension of the vertex operator  
corresponding to the coupling $g^i$. Note that $\Delta_i - 2$ is the
anomalous dimension of this coupling under quantum (world-sheet) corrections.  

Let us now consider the $\Lambda$-Universe as a \emph{non-conformal} 
gravitational background of a string~\cite{emn}. 
The background is non-conformal 
since to ${\cal O}(\alpha ')$ the relevant Einstein equation is:
\begin{eqnarray} 
\alpha ' R_{\mu\nu} =(\alpha')^2\Lambda g_{\mu\nu} \ne 0~; \qquad \Lambda > 0~,
\label{einstcosmol}
\end{eqnarray}
from which we observe that the relevant $\sigma$-model 
Weyl anomaly coefficient $\beta^G$ is non zero (c.f. (\ref{ricci})).

The conventional framework in string theory is to interpret such 
non-vanishing contributions  to the $\beta$-functions as arising from 
higher world-sheet topologies, e.g. \emph{dilaton tadpoles}, which contribute
to the ultraviolet world-sheet divergences and thus need regularization.
Such a regularization can be taken care of by adding appropriate counterterms
at a tree level in the $\sigma$-model, which are such so as to cancel the 
higher-genus infinities~\cite{fs}.  

The problem with such an approach is the convergence of the
higher-genus surface resummation, as well as the fact that 
if this were the mechanism for generating a cosmological 
constant in string theory, the resulting value should be expected
on generic grounds to be of order Planck (in appropriate units).
It would be very hard to reconcile such a mechanism with the smallness
of the currently observed cosmological constant~\cite{wmap}. 

In \cite{emn} we have proposed an alternative approach, which allows
for \emph{relaxation} mechanisms to be employed in string theory, based 
on the above-described Liouville string approach. According to the latter,
we may view the $\Lambda$-Universe as a non-critical $\sigma$-model 
which needs Liouville dressing. The origin of the non-criticality in this 
approach is an issue that can be dealt with 
in the context of detailed models~\cite{diamand,gm}, 
which we shall not describe here.
For our purposes we only mention that 
departure from conformality may be induced, for instance,
by cosmically \emph{catastrophic} events, such as the collision
of two \emph{brane} worlds~\cite{gm}. 
The non-critical $\sigma$-model in such a case 
represents stringy matter excitations on the brane corresponding to  
the observable world, and the initial central charge deficit 
is nothing other than the effective potential between the two branes
when they are closed to each other (soon after the collision). The latter
can then be computed by perturbative methods 
via the exchange of pairs
of open strings stretched between the branes.

We note at this stage that the advantage of this approach is that now
higher genus world-sheet or other effects may generate Planck size
cosmological constant, which, however, upon appropriate Liouville
dressing results in \emph{relaxing} to zero vacuum energy density,
once the Liouville mode is \emph{identified}
with the cosmological \emph{target time}~\cite{emn}. 

This identification constraints the dynamics of the 
$(D+1)$-dimensional space time on a D-dimensional 
hypersurface. It is important to notice that 
in cases of physical interest~\cite{gm} 
such a constraint is obtained \emph{dynamically}, as corresponding to 
\emph{minimization} of an appropriate effective potential in the 
$D$-dimensional target space.

Upon this identification, one can show~\cite{emn} that the time evolution 
of a matter density matrix $\rho_m$, 
propagating in such non-critical string backgrounds,has the form:
\begin{eqnarray} 
{\dot \rho}_m(\lambda, t) = i[\rho_m, {\cal H}] + : \beta^i(\lambda) 
{\cal G}_{ij} [\lambda^j, \rho_m ] :
\label{liouvevoldm}
\end{eqnarray}
where ${\cal H}$ is the matter low-energy Hamiltonian, and the 
$: \dots :$ denote appropriate normal ordering of the quantum operators
$\lambda^i$~\footnote{In $\sigma$-model perturbation theory, target-space 
\emph{canonical} quantization
is achieved formally upon summing up world-sheet topologies~\cite{emn}.
It can be shown that the Helmholtz conditions for canonical quantization
in Liouville $\sigma$-models are preserved upon the identification 
of time with the Liouville mode.}. In this sense, one may view the various 
partitions of the operator $\beta^i{\cal G}_{ij}$, which is expanded in 
powers of $\lambda^i$, as expressing the various Lindblad 
``environmental'' operators for the Liouville string. The presence of the 
environment is manifested through the non-vanishing $\beta^i \ne 0$, 
and is attributed to the fact that the set of the background couplings $g^i$
considered in the deformed $\sigma$-model at hand is \emph{not a complete one}. 
In concrete examples~\cite{emn,diamand,gm} including black hole 
and other singular backgrounds in string theory, such as colliding branes, 
one may indeed identify $\sigma$-models in which \emph{exactly marginal} 
deformation operators may be constructed, which however involve
either \emph{non-propagating} or, in general, \emph{inaccessible to local scattering experiments} 
solitonic gravitational degrees of freedom. Truncating the theory to 
the local degrees of freedom accessible to low-energy experiments, then, 
defines an \emph{effective non-critical} (Liouville) string theory.

\begin{figure}[t]
\begin{center}
\includegraphics[width=.3\textwidth]{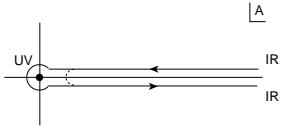}
\end{center}
\caption[]{Steepest-descent contour for integrating the Liouville
zero mode $\varphi$ in a non-critical $\sigma$-model. The curve lies 
in the
complexified world-sheet area $A$ plane.
Upon the identification
of the Liouville mode with the target time this contour becomes 
similar to closed time like paths in open (non-equilibrium) field theories,
and leads to non factorisable \$-matrices, $\$ \ne SS^\dagger$,
due to world-sheet ultraviolet
infinities (in the region $A \to 0$).}
\label{steepest}
\end{figure}

The presence of the non-Hamiltonian terms in (\ref{liouvevoldm}), 
proportional to $\beta^i \ne 0$, implies \emph{breaking} of CPT symmetry
in target space, due to the fact that, in general, 
a proper S-matrix cannot be defined 
in Liouville strings~\cite{emn}. This stems from a steepest-descent 
contour over which one defines the Liouville-mode path integral in a $\sigma$-model~\cite{emn,kogan} (c.f. figure \ref{steepest}). 
The non-factorisability of the \$-matrix, $\$ \ne SS^\dagger$, 
and hence the ill-definition of a proper scattering amplitude, 
in such cases is due to ultraviolet divergences in the small-world-sheet-area
limit~\cite{emn}. On the other hand, in the infrared limit, $A \to \infty$, 
the critical string situation, with a well-defined S-matrix, is recovered. 
The above observations follow formally from considering 
infinitesimal world-sheet 
Weyl transformations of correlation functions of vertex 
operators~\cite{emn}. Let an N-point correlator, 
${\cal A}_N = <V_1 \dots V_N>$, of a Liouville $\sigma$-model, on a world-sheet
with metric $\gamma_{\alpha\beta}$, $\alpha,\beta=1,2$, 
and consider an infinitesimal Weyl transformation $\gamma_{\alpha\beta}
\to (1 + \sigma (z,{\bar z}))\gamma_{\alpha\beta}$, $|\sigma| << 1$.
After some straightforward computations one obtains~\cite{emn}:
\begin{eqnarray} 
\delta_\sigma {\cal A}_N \sim \left(\sum_i \Delta_i + \frac{1}{A}\sum_i (\frac{\alpha_i}{\alpha} + \frac{Q}{\alpha})\right){\cal A}_N \sigma
\label{weylshift}
\end{eqnarray} 
where $A$ is the world-sheet area (in a `fixed-A' $\sigma$-model 
formalism~\cite{ddk}), $Q$  is the central charge deficit
of the non-critical string, and the $\alpha$'s are the gravitational
anomalous dimensions of the various couplings corresponding to 
the vertex operators $V_i$. Notice that the first term on the right hand side of (\ref{weylshift}) is independent of the area $A$, and transforms 
covariantly under Weyl shifts. This is the term that survives the 
critical string limit. On the other hand, the second set of terms 
transform \emph{anomalously}, proportionally to the sum 
of the Liouville anomalous dimensions and the central charge deficit, 
and hence are \emph{exclusive} to the Liouville nature of the $\sigma$-model.
Such terms \emph{vanish} only in the critical case, or in the \emph{infrared} 
limit $A \to \infty$, where the non-critical string approaches 
a critical-string \emph{equilibrium} situation~\cite{emn}.
Notice also that due to the presence of these $1/A$-terms, there are ultraviolet world-sheet divergences in the region $A \to 0$, which, 
result in an ill-defined scattering amplitude. In a critical string theory
such $A$-dependent terms are absent, and hence the correlator 
has a well-defined meaning 
as a scattering amplitude. On the other hand, if one defines
the Liouville-mode-path-integrated correlator 
${\cal A}_N$ on the regularizing contour of fig. \ref{steepest},
such divergences are regularized, but, as mentioned earlier, 
the resulting expression is identified with a \$-matrix element,
rather than a scattering amplitude~\cite{emn}. 
In this context, the world-sheet divergences 
may thus be held responsible for the \emph{non factorisability} 
of the \$-matrix 
to a product of proper scattering amplitudes. 

The lack of this property, implies, then, according to the arguments
of \cite{wald} reviewed above, a violation of CPT symmetry, at least 
in its 
strong form. From (\ref{liouvevoldm}) 
we observe that this violation if proportional to $\beta^i$. 
In the cosmological case of the $\Lambda$-Universe under 
consideration the non-criticality
is determined primarily by the graviton $\beta$-function (\ref{ricci}), which
in turn implies that the CPT-breaking terms in the 
evolution equation (\ref{liouvevoldm}) are of order (in, say, GeV):
\begin{eqnarray} 
 {\rm Cosmological~CPT-breaking~in~Liouville~string} = 
{\cal O}\left((\alpha ')^{3/2}\Lambda\right)
\label{liouvcpt}
\end{eqnarray} 
If one 
identifies $\ell_s$ with $1/M_P$, 
with $M_P \sim 10^{19}$ GeV, the four-dimensional Planck scale, 
and takes into account the current observational limits of $\Lambda \le
10^{-123} M_P^4$, 
then the cosmological CPT violating terms in (\ref{liouvcpt}) 
are of order less than $10^{-123} M_P \simeq 10^{-104}$ GeV. This is too small
to be detected in current or immediate future particle physics experiments,
such as neutral kaon facilities~\cite{cplear}, which are considered as 
sensitive probes of tests of 
quantum mechanics~\cite{ehns,lopez}.
However, from our considerations above, 
it follows that astrophysical 
observations, either through supernovae~\cite{snIa} or through 
CMB precision observations~\cite{wmap}, 
exhibit sensitivity that reach such small scales indirectly, 
and hence once 
the presence of a cosmological constant is confirmed 
in the future, this may be considered also as an indirect observation of an induced 
cosmological CPT violation, in the sense defined above. 

\section{Phenomenology of CPT-Violation due to Unitarity breaking} 

In general, within a quantum-gravity (QG) context, 
the Lindblad type violation of CPT implied by evolution equations
of the form (\ref{lindblad}), or (\ref{liouvevoldm}) may be tested in 
particle physics neutral meson experiments such as neutral 
kaons~\cite{ehns,lopez}, given that quantum gravity induced \emph{local} 
violations 
of unitarity, due to processes involving fluctuating microscopic
black holes (c.f. figure \ref{bhfoam}) may be much larger than
the above-mentioned global effects (\ref{liouvcpt}). This is due to the 
fact that such local effects do not necessarily 
have the time-attenuation factor
that characterizes the above-mentioned
relaxation models of the cosmological constant.
For instance, consider the case of a string propagating in a
Black Hole background. As explained in \cite{emn}, 
in this case the violation of world-sheet 
conformal invariance 
is due to interactions with the foam, as a result of the truncation 
of the set of background fields to propagating degrees of freedom only. 

In other words, the coefficients $C^i_{i_1 \dots i_n}$ in (\ref{weylcoeff}),
which would 
vanish identically if the set of couplings were complete
(exactly marginal deformations), are now  different from zero 
by terms corresponding to  
massive string states (with masses which 
are  multiples of $M_s$, the string mass 
scale). The latter are either 
non-propagating d.o.f. or solitonic states inaccessible
in principle to local scattering experiments, such as global Bohm-Aharonov
phases of matter wave functions induced by the quantum gravity
singular metric fluctuations~\cite{emn}.
In this case one can expect 
$C^i_{i_1 \dots i_n}$ to be expanded in a power series of $\alpha ' k^2$,
where $k^2$ is an (invariant) four-momentum scale, since only closed 
string scattering amplitudes correspond to gravitational degrees of 
freedom~\cite{tseytlin}~\footnote{If open strings are involved, 
the power series corresponds to gauge excitations, and 
the expansion is in terms of 
$\sqrt{\alpha ' }|k|$.}. 
Unless prevented by 
special reasons, which are not expected in general, the series  
starts from $\alpha' k^2 = k^2/M_s^2$, which in a Lorentz invariant theory 
is of order $m^2/M_s^2$, where $m$ a characteristic (rest) mass scale in the 
problem. This is the order of the $\beta$-functions appearing in 
(\ref{liouvevoldm}), and hence in such a case one would have: 
\begin{eqnarray} 
 {\rm QG-induced~CPT-breaking~in~Liouville~string} =
{\cal O}\left( \frac{m^2}{M_s} \right)
\label{liouvqgcpt}
\end{eqnarray} 
As mentioned above, the string mass scale 
$M_s$ may or may not be equal to the four-dimensional 
Planck mass $M_P \sim 10^{19}$ GeV.
Thus, if such a situation is actually encountered in nature, 
the associated CPT breaking effects are much larger than the cosmological 
effects (\ref{liouvcpt}). This is due to the fact that in these examples
the conformal invariance 
is violated at tree level in a world-sheet genus
expansion, due to a truncation of the 
spectrum. If the violation occurs at higher genera, 
as could be the case of a more conventional violation due to
graviton loops in a field-theory context, 
then one expects in that case 
a further suppression of (\ref{liouvqgcpt}) 
by an extra factor $m/M_s$, i.e. one has CPT-violating 
terms of order ${\cal O}(m^3/M_s^2)$. 

Such effects may then be bounded (or tested!) experimentally 
by taking into account that 
their presence may induce decoherence and oscillations 
in, say, neutral mesons~\cite{ehns,lopez,huet,benatti}, 
neutrinos~\cite{lisi,ben2} \emph{etc.}. 
We shall discuss first 
the neutral meson case, concentrating on the most sensitive 
probe that of neutral kaons. 

The QG induced oscillations 
are between Kaon and its antiparticle 
$K^0 \to {\overline K}^0$~\cite{ehns,lopez}.  
The modified evolution equation 
for the respective density matrices of neutral kaon matter 
in the \emph{linearized} approximation,
(\ref{lindblad}) or 
(\ref{liouvevoldm}), 
can be parametrized as follows~\cite{ehns}: 
$$\partial_t \rho = i[\rho, H] + \delta\H \rho~,$$ 
where 
$$H_{\alpha\beta}=\left( \begin{array}{cccc}  - \Gamma & -\coeff{1}{2}\delta \Gamma
& -{\rm Im} \Gamma _{12} & -{\rm Re}\Gamma _{12} \\
 - \coeff{1}{2}\delta \Gamma
  & -\Gamma & - 2{\rm Re}M_{12}&  -2{\rm Im} M_{12} \\
 - {\rm Im} \Gamma_{12} &  2{\rm Re}M_{12} & -\Gamma & -\delta M    \\
 -{\rm Re}\Gamma _{12} & -2{\rm Im} M_{12} & \delta M   & -\Gamma
\end{array}\right) $$ and 
$$ {\delta\H}_{\alpha\beta} =\left( \begin{array}{cccc}
 0  &  0 & 0 & 0 \\
 0  &  0 & 0 & 0 \\
 0  &  0 & -2\alpha  & -2\beta \\
 0  &  0 & -2\beta & -2\gamma \end{array}\right)~.$$
Positivity of $\rho$ requires:
$\alpha, \gamma  > 0,\quad \alpha\gamma>\beta^2$.
Notice that 
$\alpha,\beta,\gamma$ violate CPT, as they do not commute
with a CPT operator $\Theta$
connecting $K^0$ to ${\overline K}^0$~\cite{lopez}:  
$\Theta = \sigma_3 \cos\theta + 
\sigma_2 \sin\theta$,$~~~~~[\delta\H_{\alpha\beta}, \Theta ] \ne 0$.

An important remark is now in order. 
We should distinguish two types of CPT violation (CPTV):
(i) CPTV within a Quantum Mechanical formalism:
$\delta M= m_{K^0} - m_{{\overline K}^0}$, $\delta \Gamma = \Gamma_{K^0}-
\Gamma_{{\overline K}^0} $. 
This could be  due to (spontaneous) Lorentz violation 
and/or violations of locality (c.f. below).\\
(ii) CPTV through decoherence $\alpha,\beta,\gamma$ 
(entanglement with QG `environment', leading to modified 
evolution for $\rho$ (\ref{evolution}) and  $\$ \ne S~S^\dagger $). 

\begin{table}[thb]
\begin{center}
\begin{tabular}{lcc}
\underline{Process}&QMV&QM\\
$A_{2\pi}$&$\not=$&$\not=$\\
$A_{3\pi}$&$\not=$&$\not=$\\
$A_{\rm T}$&$\not=$&$=$\\
$A_{\rm CPT}$&$=$&$\not=$\\
$A_{\Delta m}$&$\not=$&$=$\\
$\zeta$&$\not=$&$=$
\end{tabular}
\caption{Qualitative comparison of predictions for various observables
in CPT-violating theories beyond (QMV) and within (QM) quantum mechanics.
Predictions either differ ($\not=$) or agree ($=$) with the results obtained
in conventional quantum-mechanical CP violation. Note that these frameworks can
be qualitatively distinguished via their predictions for $A_{\rm T}$, $A_{\rm
CPT}$, $A_{\Delta m}$, and $\zeta$.}
\label{Table2}
\end{center}
\hrule
\end{table}

The important point is that the two types of CPTV can 
be \emph{disentangled experimentally}~\cite{lopez}. 
The relevant observables are defined as $ \VEV{O_i}= {\rm Tr}\,[O_i\rho] $.
For neutral kaons, one looks at decay asymmetries 
for $K^0, {\overline K}^0$, defined as: 
$$A (t) = \frac{
    R({\bar K}^0_{t=0} \rightarrow
{\bar f} ) -
    R(K^0_{t=0} \rightarrow
f ) }
{ R({\bar K}^0_{t=0} \rightarrow
{\bar f} ) +
    R(K^0_{t=0} \rightarrow
f ) }~,$$ 
where $R(K^0\rightarrow f) \equiv \Tr[O_{f}\rho (t)]=$ denotes the decay rate
into the final state $f$ (starting from a pure $ K^0$ state at $t=0$).

In the case of neutral kaons, one may consider the 
following set of asymmetries:
(i) {\it identical final states}: 
$f={\bar f} = 2\pi $: $A_{2\pi}~,~A_{3\pi}$,
(ii) {\it semileptonic} : $A_T$
(final states $f=\pi^+l^-\bar\nu\ \not=\ \bar f=\pi^-l^+\nu$), $A_{CPT}$ (${\overline f}=\pi^+l^-\bar\nu ,~ f=\pi^-l^+\nu$), 
$A_{\Delta m}$.  Typically, for instance when final states are $2\pi$,
one has  a time evolution of the decay rate $R_{2\pi}$: 
$ R_{2\pi}(t)=c_S\, e^{-\Gamma_S t}+c_L\, e^{-\Gamma_L t}
+ 2c_I\, e^{-\Gamma t}\cos(\Delta mt-\phi)$, where 
$S$=short-lived, $L$=long-lived, $I$=interference term, 
$\Delta m = m_L - m_S$, $\Gamma =\frac{1}{2}(\Gamma_S + \Gamma_L)$. 
One may define the \emph{Decoherence Parameter}
$\zeta=1-{c_I\over\sqrt{c_Sc_L}}$, as a measure 
of quantum decoherence induced in the system. 
For larger sensitivities one can look 
at this parameter in the presence of a 
regenerator~\cite{lopez}. 
In our decoherence scenario, $\zeta$ depends primarily on $\beta$,
hence the best bounds on $\beta$ can be placed by 
implementing a regenerator~\cite{lopez}.

\begin{figure}[t]
\begin{center}
\includegraphics[width=.5\textwidth]{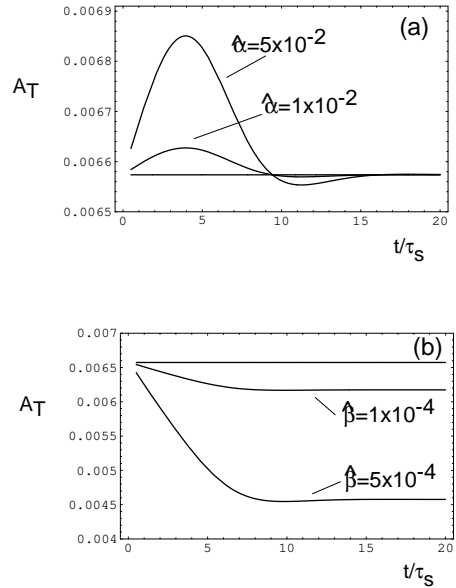}
\end{center}
\caption[]{A typical neutral kaon decay asymmetry $A_T$~\cite{lopez} 
indicating the 
effects of quantum-gravity induced decoherence.} 
\label{AT}
\end{figure}

The experimental tests (decay asymmetries) 
that can be performed in order to disentangle
decoherence from quantum mechanical CPT violating effects 
are summarized in table \ref{Table2}. 
In figure \ref{AT} we give a typical profile of a decay asymmetry,
that of $A_{T}$~\cite{lopez},
from where bounds on QG decoherening parameters can be extracted. 
The other asymmetries may be studied in a similar fashion.
Details can be given in \cite{lopez}, where we refer the interested
reader for details. 
Experimentally, the best available bounds at present come from 
CPLEAR measurements~\cite{cplear} 
$\alpha < 4.0 \times 10^{-17} ~{\rm GeV}~, ~|\beta | < 2.3. \times
10^{-19} ~{\rm GeV}~, ~\gamma < 3.7 \times 10^{-21} ~{\rm GeV} $,
which are not much different from 
theoretically expected values 
$\alpha~,\beta~,\gamma = O(\xi \frac{m_K^2}{M_{P}})$, where $m_K$ 
is the Kaon rest mass.

One may extend the above formalism to study correlated Kaon states,
as those produced in a $\phi$ decay~\cite{huet},
$\phi \to K^0 {\overline K}^0$. It is interesting to note
that in such cases the non-quantum mechanical terms in (\ref{lindblad}) 
produce terms that apparently violate energy and angular momentum,
at a microscopic level,  
and this is consistent with generic properties of the formalism~\cite{lopez}.
It must be stressed though that the formalism 
remains still an open issue, given that the evolution of correlated states
may require genuine two-particle state variables for 
the non-quantum mechanical parts,  whilst in \cite{huet} one used only 
single-particle variables $\alpha,\beta,\gamma$. An additional 
point is the validity of the requirement of \emph{complete positivity} 
of the reduced 
density matrix of 
the correlated Kaon states~\cite{benatti}, 
which would impose further restrictions on the set of the QG decoherence 
parameters. However, 
in view of potential \emph{non-linearities} of quantum gravity, or
other mean-field effects, it is not clear whether such a requirement 
actually holds~\cite{chachor,emnadler}. 
Such issues present interesting challenges for the theory.

Finally, when the above-described 
formalism of open quantum 
systems~\cite{ehns} is applied to neutrinos,
induced oscillations between neutrino flavours may occur as a result
of decoherence~\cite{lisi,ben2}, which may be 
independent of neutrino masses. Fitting currently available data from 
atmospheric and solar neutrinos~\cite{lisi} one may obtain sensitivities
which exceed the Planck scale by far. For instance, concentrating on the 
QG-decoherence-induced transition probability $P_{\nu_e \to \nu_\mu}(t)$,
and \emph{assuming} that the dissipative environmental (QG) effects
are of order 
\begin{eqnarray} 
\frac{<E>^2}{M_{QG}}~, 
\label{optimistic}
\end{eqnarray} 
where $<E>$ is an average neutrino energy,
the authors of \cite{ben2} estimated the decoherence 
effects to be smaller than $10^{-27}$ GeV, implying an 
effective QG scale for massless 
neutrino much larger than $M_P$ for $10^3$ GeV energies. 
On the other hand, an earlier analysis~\cite{lisi} 
by means of the super-Kamiokande data~\cite{kamioka} for
$\nu_\mu - \nu_\tau$ oscillations yields a somewhat weaker 
bound for the QG-decoherence effects $< 3.5 \times 10^{-23}$ GeV 
in the massive neutrino case,  
using the super Kamiokande value~\cite{kamioka} 
for the neutrino squared mass difference 
$\Delta m^2 = m_2^2 - m_1^2 = 3 \times 10^{-3}$ eV$^2$ 
between the oscillating states.   
In the massless neutrino case, the analysis of \cite{lisi}
yields more stringent bounds for 
the decoherence parameter(s), of order $10^{-27}$ GeV, as in \cite{ben2}. 

However, theoretically, for massive neutrino models of decoherence, 
more conservative estimates  
have been presented for \emph{some models}~\cite{adler} 
according to which the decoherence parameters are of order 
\begin{eqnarray} 
\frac{(\Delta m^2)^2}{<E>^2M_{QG}}
\label{pessim}
\end{eqnarray}  
which is much more pessimistic than (\ref{optimistic}), and 
is probably not detectable at immediate future facilities. 
The situation is far from being conclusive, however, as it depends
on the details of the QG environment and its interaction with the 
subsystem~\cite{emnadler}.

A comment we would like to make concerns the presence of energies
$E$ in both estimates (\ref{optimistic}) and (\ref{pessim}), 
which makes them appropriate only for \emph{non Lorentz invariant} 
formulations. For Lorentz Invariant (LI)
cases one would expect only rest mass terms
to appear, for instance (\ref{pessim}) 
could be replaced by $\left( \Delta m^2/M_{QG} \right)$ 
in minimal suppression
models \emph{etc.} For \emph{massless} neutrinos the LI
formulation is tricky, as one lacks a fundamental low-energy mass scale. 
Some steps towards LI decoherence are taken in \cite{lidecoh}, 
where a LI decoherence has been defined in terms of \emph{intrinsic 
quantum mechanical uncertainties} of spatial 
translations, like the ones entering the generalized uncertainty principle
in string theory~\cite{venezia}. 
However,  
in our opinion, the 
issue remains a challenging open one.

\section{Quantum Gravity, Lorentz violation and CPT} 

The above discussion assumed that Lorentz invariance is maintained
by quantum gravity. This may not be the case in certain backgrounds, for 
instance those 
involving \emph{spontaneous breaking of Lorentz symmetry}
by means of certain tensor quantities acquiring vacuum expectation 
values~\cite{kostel}, $<A_{\mu_1 \dots \mu_N}> \ne 0$, 
which may characterize some string theory backgrounds. Another set of models
with potential Lorentz violations
are 
loop quantum gravity models~\cite{loopgrav}, as well as 
approaches to quantum gravity viewing the Planck 
length as a `real length'~\cite{amelino}, 
for which the requirement of not being subjected
to Lorentz contraction leads to modified dispersion relations for 
matter probes, including photons,
already in flat Minkowski space times.
The breaking of Lorentz symmetry violates the requirement (ii) of the CPT theorem stated in the introduction, and hence leads to 
a breaking of CPT symmetry of the associated field theory. 

In such models one may encounter situations in which the violations of Lorentz
Invariance (LIV) and, hence, CPT, are \emph{minimally} suppressed
by a single power of the Planck mass 
scale, 
$M_P\sim 10^{19}$~GeV  
(or, in general, a QG scale,  
$M_{QG}$, which may be different from $M_P$), and are typically of order 
\begin{eqnarray} 
{\rm Minimally~suppressed~LIV~CPTV} = {\cal O}\left(\frac{E^2}{M_P}\right) 
\label{livcptv}
\end{eqnarray}
where $E$ is a typical energy scale of the matter probe, in the frame of 
observation. 

We must note here that as a result of LIV there is a \emph{frame
dependence} of the results, and probably a preferred frame, which 
may be taken to be the CMB frame (but this may not the only possibility).

The basic formalism is described in \cite{kostel}, where one considers
\emph{modified Dirac equation (MDE)} for fermions
in the so-called Standard Model Extension (SME).
In view of the recent `massive' production of antihydrogen 
(${\overline H}$ ) at 
CERN \cite{cernhbar}, 
which implies that 
interesting direct tests of CPT invariance using ${\overline H}$
are to be expected in the near future, we   
consider for our purposes here the specific case of MDE 
for Hydrogen $H$ (anti-hydrogen ${\overline H}$), although the 
formalism is generic. 
Let
the spinor $\psi$  represent the electron  
(positron) with charge $q=-|e| (q=|e|)$ 
around a proton (antiproton) 
of charge $-q$. Then the MDE reads:
$$\left( i\gamma^\mu D^\mu - M -  
a_\mu \gamma^\mu - b_\mu \gamma_5 \gamma^\mu -
\frac{1}{2}H_{\mu\nu}\sigma ^{\mu\nu}
+  ic_{\mu\nu}\gamma^\mu D^\nu + id_{\mu\nu}\gamma_5\gamma^\mu D^\nu  
\right)\psi =0,$$
where $D_\mu = \partial_\mu - q A_\mu$, $A_\mu = (-q/4\pi r, 0)$ Coulomb 
potential. The parameters $a_\mu~, b_\mu $ induce 
CPT and Lorentz violation, while the parameters $c_{\mu\nu}, d_{\mu\nu},  
H_{\mu\nu} $ induce Lorentz violation only.

In SME models there are energy shifts between states 
$|J,I;m_J,m_I>$, with $J (I)$ denoting electronic (nuclear) 
angular momenta. Using 
perturbation theory, one finds~\cite{kostel}:
\begin{eqnarray} 
&&\Delta E^H (m_J, m_I) \simeq  a_0^e + a_0^p - c_{00}^e m_e - c_{00}^p m_p 
+ (-b_3^e + d_{30}^em_e + H_{12}^e)\frac{m_J}{|m_J|} + \nonumber \\
&&(-b_3^p + d_{30}^p m_p + H_{12}^p)\frac{m_I}{|m_I|}~, \nonumber 
\end{eqnarray} 
where $e$ electron; $p$ proton. The 
corresponding results for antihydrogen (${\overline H}$) 
are obtained upon:
$$a_\mu^{e,p} \rightarrow -a_\mu^{e,p}~,~ 
b_\mu^{e,p} \rightarrow -b_\mu^{e,p}~,~ 
d_{\mu\nu}^{e,p} \rightarrow d_{\mu\nu}^{e,p}~, 
H_{\mu\nu}^{e,p} \rightarrow H_{\mu\nu}^{e,p}~.$$

One may study the spectroscopy of \emph{forbidden transitions 1S-2S}: 
If CPT and Lorentz violating parameters are
constant they drop out to leading order energy shifts in free H 
(${\overline H})$.  Subleading effects are then suppressed by 
the square of the fine structure constant:
$\alpha^2 \sim 5 \times 10^{-5}$, specifically:  
$\delta \nu^H_{1S-2S} \simeq -\frac{\alpha^2b_3^e}{8\pi}$.
This is too small to be seen. 

But what about 
the case where atoms of $H$ (or ${\overline H}$) are 
in magnetic traps? Magnetic fields 
induce hyperfine Zeeman splittings in 1S, 2S states.
There are four spin states, mixed under the 
the magnetic field B ($|m_J,m_I>$ basis):
$|d>_n =|\frac{1}{2},\frac{1}{2}>$,
$|c>_n ={\rm sin}\theta_n|-\frac{1}{2},\frac{1}{2}> +
{\rm cos}\theta_n |\frac{1}{2},-\frac{1}{2}>$,
$|b>_n =|-\frac{1}{2},-\frac{1}{2}>$,
$|a>_n ={\rm cos}\theta_n|-\frac{1}{2},\frac{1}{2}> -
{\rm sin}\theta_n |\frac{1}{2},-\frac{1}{2}>$,  
where ${\rm tan}2\theta_n= (51 {\rm mT})/{\rm n^3B}$. The 
$|c>_1 \to |c>_2$ transitions yield dominant effects for 
CPTV~\cite{kostel}: 
\begin{eqnarray} 
&& \delta \nu_c^H \simeq -
\frac{\kappa (b_3^e-b_3^p-d_{30}^em_e + d_{30}^pm_p-H_{12}^e + 
H_{12}^p)}{2\pi}~, \nonumber \\  
&&\delta \nu_c^{\overline H} \simeq 
-\frac{\kappa (-b_3^e+b_3^p-d_{30}^em_e - d_{30}^pm_p-H_{12}^e + H_{12}^p)}{2\pi}~, \nonumber \\
&&\Delta \nu_{1S-2S,c} \equiv \delta \nu_c^H -\delta \nu_c^{\overline H} \simeq
-\frac{\kappa(b_3^e-b_3^p)}{\pi}~, \nonumber 
\end{eqnarray}  
where $\kappa ={\rm cos}2\theta_2 - {\rm cos}2\theta_1$,  
$\kappa \simeq 0.67 $ for $B=0.011$ T. 
Notice that $\Delta \nu_{c\to d} \simeq -2b_3^p/\pi~$, and,  
\emph{if a frequency resolution of 1 mHz is attained}, one may obtain 
a bound $|b_3| \le 10^{-27} {\rm GeV}~$.

\begin{figure}[t]
\begin{center}
\includegraphics[width=.8\textwidth]{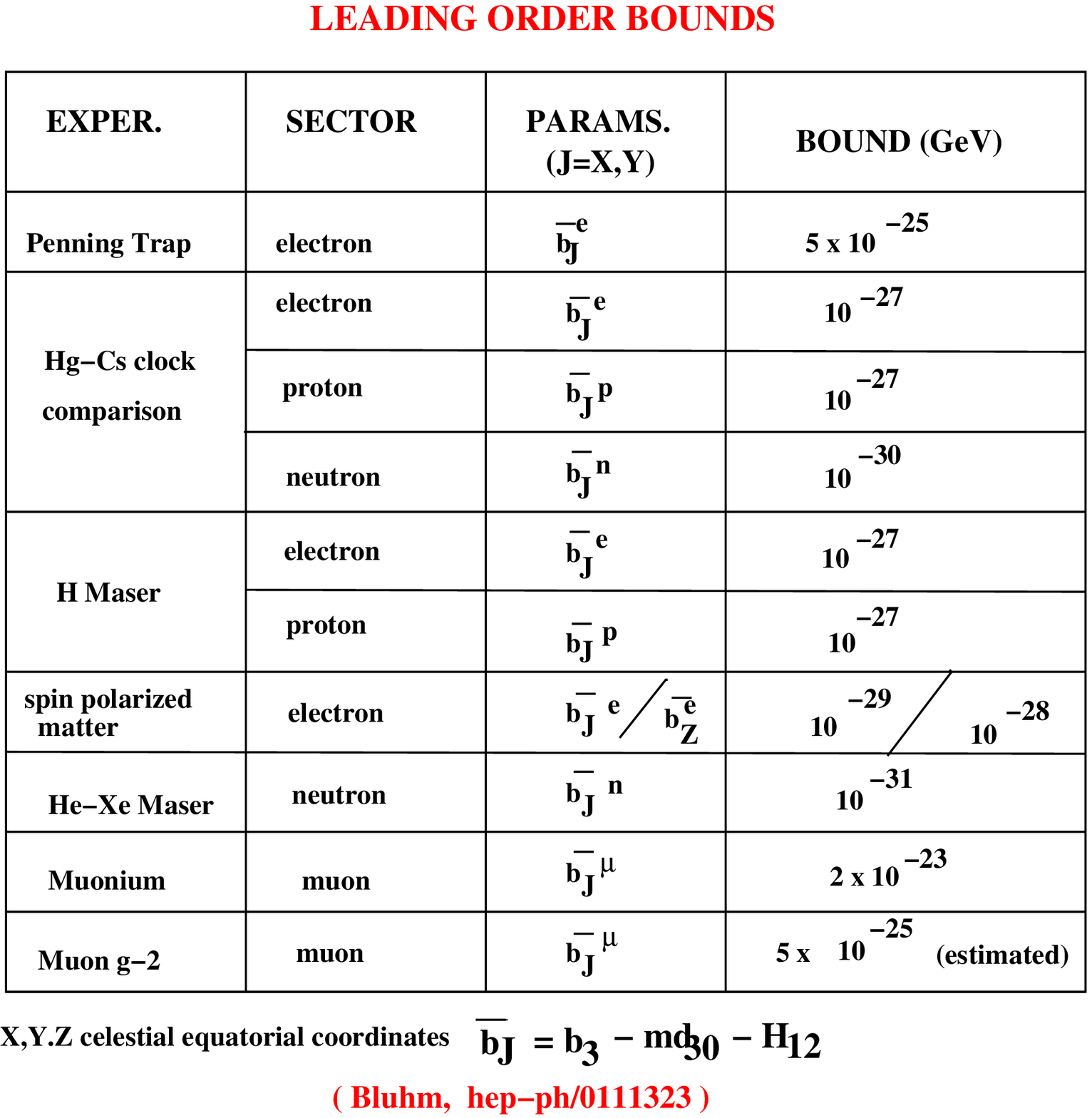}
\end{center}
\caption[]{Table summarizing recent bounds of CPT violating 
parameter $b$ in the Standard Model extension 
from atomic and nuclear physics spectroscopic tests~\cite{bluhm}.}
\label{bluhm}
\end{figure}

The existence of a preferred frame implies very stringent restrictions
on such LIV and CPTV terms, most of which stem from observations (both terrestrial and astrophysical) pertaining to \emph{electrically charged} 
fermions~\cite{livfermions,synchro}. These observations seem to exclude 
QG-induced \emph{minimally suppressed (linear)} 
modifications of matter dispersion relations, given that the sensitivity 
of such experiments exceeds the Planck scale $10^{19}$ GeV 
by several orders of magnitude, 
and hence such models are excluded on naturalness grounds. 

For instance, one of the most stringent constraints at present on 
linearly modified electron dispersion relations 
is obtained from observations of the synchrotron radiation from Crab 
Nebula~\cite{synchro}, with sensitivity that exceeds 
$M_P \sim 10^{19}$ GeV by nine orders of magnitude.  
Also, one may obtain high sensitivities in atomic and nuclear
physics experiments~\cite{livfermions}. A summary 
of the various sensitivities from atomic physics 
experiments is given in the table of figure 
\ref{bluhm}~\cite{bluhm}, where we can see that in some cases
CPT and LIV violating terms for electrons (or electrically charged fermions)
may reach sensitivities up to $10^{-31}$ GeV ! 

\section{Non-Critical String 
Models of Foam, the Equivalence principle and the Evasion (?) of the Constraints} 

Although from the above discussion it seems that current experiments exhibit sensitivities that exclude linearly modified dispersion relations, however 
the actual situation may be more subtle. In fact, we shall argue below that 
in a Liouville string approach to foam, there is a \emph{violation} 
of the \emph{equivalence principle} in 
the sense that not all matter probes
interact the same way with the gravitational foamy backgrounds. 
In particular, photons, or at most particles \emph{neutral} 
under the (unbroken) standard 
model gauge group $SU(3)_c \otimes U(1)$, are the only types 
that may exhibit modified dispersion relations~\cite{ems}. 
The rest of the low-energy modes are insensitive to the 
quantum gravity foamy effects, as far as dispersion 
relations are concerned. 

A concrete model of Liouville string foam 
with this property 
has been presented 
in \cite{emnliouvfoam}. According to this model our world 
is a three brane, and the observable matter particles (including radiation) 
are viewed as open string excitations on it, with their ends attached 
to the brane. The brane is embedded in a higher (bulk) space time 
in which only closed strings (gravitational d.o.f.) are allowed to propagate.
The foam is obtained by assuming quantum fluctuations of the brane world
which are such that (virtual) D-particle defects can emerge from the brane.
These are short-lived excitations, with average life time 
$\tau \sim \frac{\ell_s}{c}$ (where $\ell_s$ is the string length, 
and $c$ the speed of light in empty space). The D-particles are 
degenerate forms (point-like) of D-branes~\cite{polchinski}, but as their 
higher-dimensional counterparts they can \emph{capture} open strings, 
since the latter may have their ends attached to the D-particle.  
From a conformal field theory point of view the capture stage 
may be described by the following $\sigma$-model deformations~\cite{recoil} 
\begin{eqnarray} 
\oint _{\partial \Sigma} u_i X^0 \Theta_\epsilon (X^0)\partial_n X^i~, 
\quad \oint _{\partial \Sigma} \epsilon y_i \Theta_\epsilon (X^0)\partial_n X^i~, 
\label{logpair} 
\end{eqnarray} 
where $y_i$ denotes the location of the D-particle on the D3 brane world, 
and $u_i =g_s \frac{k_1 + k_2}{M_s}$, denotes the 
recoil velocity (proportional to the momentum transfer), with $g_s$ the string 
coupling, and $M_s$ the string mass scale. The parameter $\epsilon
\to 0^+$ regulates the Heaviside operator $\Theta (X^0)$. It can be shown that 
the pair of deformations (\ref{logpair}) closes to form a \emph{logarithmic conformal} algebra (LCFT), provided $\epsilon^{-2}$ is identified with the 
logarithm of the world-sheet RG scale. Such LCFT
lie in the border line between conformal algebras
and general two-dimensional field theories. The operators (\ref{logpair}) 
are \emph{marginally relevant} in a world-sheet RG sense, 
with anomalous dimensions $\epsilon^2$. 
This has 
important implications for the departure of the deformed $\sigma$-model
from the conformal point, hence the need for Liouville dressing~\cite{ddk}. 

This dressing results in the appearance of non trivial metric deformations
in space time, which physically are interpreted  as a back reaction 
of the recoiling D-particle, during the capture stage, onto the neighboring 
space time. The resulting metric assumes the form (asymptotically in time 
after the capture stage)~\cite{recoil}:
\begin{eqnarray} 
G_{00} = -1~, \qquad G_{ij} =\delta_{ij}~, \qquad G_{0i} = u_i 
\label{recoilmetric}
\end{eqnarray}
This change in the space time background, during every interaction
of the matter probe with the D-particle, 
causes a \emph{modification} in the dispersion relation of the probe,
as a result of the relation:
\begin{eqnarray} 
p_{\mu} p_{\nu} G^{\mu\nu} = -m^2~ \to ~ -E^2 + {\vec p}^2 + 2 E {\vec p} \cdot {\vec u} = -m^2 ( 1 + {\vec u}^2 )  
\label{moddis}
\end{eqnarray} 
It should be stressed that 
this modification is consistent with the general coordinate diffeomorphism
invariance of the target space of the stringy $\sigma$-model, which is respected by the foam. From (\ref{moddis}) it is evident that in this way 
one obtains \emph{linear} modifications in the dispersion relations, 
and hence such models, if valid for electrically charged fermions, 
would have been excluded in the sense of yielding $M_s/g_s > 10^{29}$ GeV. 
We stress that 
only the \emph{subluminal} branch of the 
modifications in the dispersion relation is consistent with the capture stage~\cite{ems}, and in fact it is only such modifications that are allowed
by the string dynamics~\cite{emnliouvfoam}.

Mathematically, of course, one could still save such models by adjusting 
$M_s$ or $g_s$ appropriately so as to meet such sensitivities, but then 
one probably faces a naturalness problem, in that one should explain 
why this particular model of foam is characterized by such small $g_s$
(assuming $M_s \sim M_P$). 

Fortunately for the model of \cite{emnliouvfoam} there seems to be 
a \emph{way out}, which in fact is based on \emph{gauge symmetry principles}. 
The point is that the capture of an open string by a D-particle has been shown 
in \cite{recoil} to describe $U(1)$ gauge excitations, whose dynamics is 
described by a 
Born-Infeld (BI) effective Lagrangian 
${\cal L}_{BI} = \sqrt{G_{\mu\nu} + \alpha ' F_{\mu\nu} }$,
with $F_{\mu\nu}$ the $U(1)$ 
Maxwell field strength, 
and $G_{\mu\nu}$ the background metric. In general, $N$ coincident $D$-particles exchanging open strings stretched between them, will transform according to the adjoint representation of the $U(N)$ group, and obey a non Abelian BI action~\cite{polchinski}. 
This includes also the 
recoil case~\cite{recoil}. Thus we conclude that the capture/recoil 
of open strings from a group of $N$ D-particle excitations in the foam 
will result in gauge excitations on the D3 brane world. There is no way 
of obtaining other excitations, transforming for instance according to 
the fundamental representation of the group. For the latter to happen one needs
\emph{intersecting} brane configurations at angles 
$\theta$~\cite{intersecting},
the latter serving to define appropriate chirality for fermionic excitations. 
D-particles are by definition parallel among themselves, and one cannot thus get 
chiral fermions by exchanging strings among them, a process characterizing 
the capture/recoil case. 

Given that in a space-time foam model the quantum numbers of the 
`vacuum' must be preserved~\cite{wheeler} by the interactions of matter with the foam d.o.f., we must conclude from the above considerations that 
only string modes which transform in the adjoint representation 
of the (unbroken) standard model group $U(3) \simeq SU(2) \otimes U(1)$, 
and are therefore gauge excitations,  can interact in the above 
sense with the D-particles and have modified dispersion relations 
of the type (\ref{moddis}).

The above result holds at \emph{tree level} in a world-sheet formalism.
From a world sheet view point the capture of the open string or closed string 
by a D-particle implies either changes in the Boundary conditions
of the open string (from Neumann (N) to Dirichlet (D)~\cite{polchinski}), 
or splitting of the closed string to a pair of open ones, with 
Dirichlet boundary conditions. 
A natural question arises in connection with higher 
world-sheet topologies, for instance
one loop (annulus) world-sheet 
graphs. Such loops may express, for instance, 
self energy
parts of electrons. Such quantum corrections involve propagating 
photons (and other (actually, an infinity of) string excitations), 
and one may expect that these will interact 
with the D-particles of the foam.
The issue as to the precise effects of these loop calculations 
on the self energy of the electrons and other charged fermions 
is under investigation. 
This may also depend on the complicated
dynamics of the (yet unknown) M-theory that describes such defects.
Such issues deserve further study before conclusions are reached.

The above considerations exclude 
electrons, and other electrically charged fermions 
(e.g. quarks) from being captured by the D-particle (at tree level at least), 
given that the capture of 
such excitations by the D-particle would lead to violations 
of electric charge~\cite{ems}. Thus, modulo 
higher order string-loop effects,
the most stringent constraints
on the linear modifications of the dispersion relations, which are based 
on such fermions~\cite{livfermions,synchro}, are evaded in the Liouville
string model of \cite{emnliouvfoam}. 
On the other hand, \emph{photons} (and probably gluons~\footnote{Confinement 
of gluons makes their interaction with the foam subtle. In this review we 
simply assume that gluons can interact like photons with the D-particles,
and do not discuss the matter further. We shall come back to such issues
in a forthcoming publication.}), \emph{can} interact with D-particles 
\emph{at tree level in $\sigma$-model perturbation theory}, 
thereby exhibiting linearly 
modified dispersion relations and \emph{subluminal} 
refractive indices (the subluminal nature is due to the Born-Infeld 
electrodynamics~\cite{emnliouvfoam}).

Such effects can be tested by $\gamma$-Ray Burst 
studies (GRB) of the arrival times of photons, as suggested in \cite{nature}. The sensitivity of such experiments, at present, is such that the QG effective 
scale for photons exhibiting linear dispersion relations (\ref{moddis}) 
is $M_{QG} > 10^{15}$ GeV~\cite{mitsou}. 
In the model of \cite{emnliouvfoam}, the interaction of a photon excitation
with a D-particle is accompanied by a \emph{random phase}~\cite{ems}
of the re-emitted photon after the capture stage. Such 
an effect also invalidates claims made in \cite{phasecoh} for 
excluding linear modifications in the photon dispersion relations~\footnote{Apart from such random phases, we also note that the analysis of \cite{phasecoh} 
overestimated the effects of the foam by a huge factor, as argued in \cite{ng}, which by itself invalidates their conclusions.}.

It is evident from 
the above considerations that \emph{details} in the \emph{dynamics}
of \emph{space-time foam} do \emph{matter} in discussing the 
pertinent phenomenology. Many conclusions, based on generic arguments
from our experience so far
with ordinary local quantum field theories, may be misleading when applied to 
quantum gravity. 

\section{CPT Breaking through Locality Violation and Neutrino Anomalies(?)}

A final topic we would like to discuss briefly is the violation of CPT 
in the neutrino sector of the theory. The reason why we chose to discuss 
this topic in a separate section is the relatively recent claims from the 
LSND experiment~\cite{lsnd} 
on evidence for observation of oscillations in the antineutrino sector 
${\overline \nu}_e \leftarrow\rightarrow {\overline \nu}_\mu$,
but the absence of such oscillation in the corresponding neutrino 
sector~\footnote{The initial 2.6$\sigma$ hint for  $\nu_\mu - 
\nu_e $ decreased to 0.6$\sigma$, while the signal for 
antineutrino oscillations remained.}. 
The situation will be clarified experimentally, 
when experiments looking directly  
for $\nu_\mu \to \nu_e$, will be in operation,  
like MiniBoonNE~\cite{miniboone}~\footnote{The material of this review 
was presented and written before the very recent announcement on
September 7 2003 by the Sadbury Neutrino Observatory (SNO), on improved
measurements of neutral current events, which, together with 
other existing data, confirm the three-neutrino scenaria. The phenomenology 
presented in this section is before these results. For a status of the 
three neutrino oscillations after the latest SNO results we refer the 
reader to \cite{valle}. The ideas on CPT violation 
presented here may then be tested by such updated data.}. 

\begin{figure}[t]
\begin{center}
\includegraphics[width=.4\textwidth]{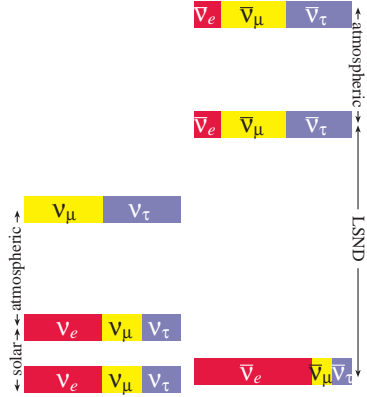}
\end{center}
\caption[]{In the scenario of \cite{murayama}, the LSND anomaly may be 
explained by a CPTV mass difference between neutrinos and 
antineutrinos.}
\label{cptvmassnu}
\end{figure}

If one associates oscillations with Dirac mass terms for neutrinos, then 
the LSND result~\cite{lsnd}, 
if correct, points towards CPT violation in the sense
of an antineutrino Dirac mass being much higher than a neutrino one, the 
mass difference between neutrino and antineutrino being of the order of 
1 eV. This idea 
was put forward first in \cite{murayama} in a 
purely phenomenological setting (c.f. fig. \ref{cptvmassnu}), 
in an attempt to explain the LSND anomaly~\cite{lsnd} without 
invoking a \emph{sterile} neutrino.

There are many microscopic theories which could lead to 
such violations. One obvious one is the spontaneous 
violation of Lorentz symmetry,
along the lines of the SME~\cite{kostel}, described briefly above, 
whose applications to the massive neutrino physics 
has been discussed recently in \cite{kostelnu}. 

On the other hand, in the 
Liouville model of D-particle foam of \cite{emnliouvfoam},
neutrinos may not have modified dispersion relations,
as their capture stage by D-particles would not seem to 
respect the gauge quantum numbers of the vacuum. 
We remind the reader that neutrinos, viewed as 
open string excitations on the brane, transform in the 
fundamental representation of appropriate gauge groups, while 
a string-D-particle `composite', describing the capture stage, 
behaves as a gauge field excitation in the model, transforming 
in the adjoint representation of the unbroken standard model 
group~\footnote{Nevertheless, one may envisage 
the possibility of neutrinos interacting with D-particle defects 
in brane models intersecting at angles~\cite{intersecting}, 
where chiral fermions are open string excitations 
localised on the intersection hypersurface, viewed as our world.
We do not know yet whether puncturing the intersection with D-particles
can be consistent with ND chiral open strings with 
mixed boundary conditions, corresponding to neutrinos, 
with one end attached to the 
D-particle (D), and the other end free (N) on the intersection, that would 
allow neutrinos (as electrically neutral) to have modified dispersion
relations.}. 
However, in other models of QG, where gravitational 
fluctuations imply the existence of an `environment'~\cite{loopgrav,lisi}
neutrinos may 
have non trivial 
refractive indices (for the massless species).
Such modifications, as we discussed in previous sections, 
may lead to 
CPTV and LIV 
effects of order $g_sE^2/M_s $, where $E$ is a typical neutrino energy
at the frame of observation. 
Such terms seem to be much smaller than the suggested 
neutrino-antineutrino
squared mass difference of ${\cal O}(0.1-1~{\rm eV}^2)$ to explain the 
LSND anomaly~\cite{lsnd,murayama}.

A more radical approach has been suggested earlier in \cite{barenboim},
according to which the neutrino sector of the standard model 
exhibits \emph{non local} interactions among the neutrinos, responsible for 
the generation of CPT Violating but Lorentz invariant mass spectra
for neutrinos, fitting phenomenologically the LSND results.  
Specifically, the model invokes a Dirac-like theory of 
neutrinos (called ``homeotic'') 
with \emph{both} positive (+) and negative (-) energies, in which the 
spinors are described by: 
\begin{eqnarray} 
&& \psi_+(x) = u_+(p)e^{-ip\cdot x}, \quad p^2 = m^2, ~p_0 > 0 \nonumber \\
&&\psi_-(x) = u_-(p)e^{-ip\cdot x}, \quad p^2 = m^2, ~p_0 < 0 \nonumber \\
&& (p_\mu \gamma^\mu - m \epsilon (p_0))u_\pm (p) =0~, 
\label{homeotic}
\end{eqnarray} 
where $\psi$ are the fermionic \emph{Dirac} neutrino fields, and  
$\epsilon (p_0)$ is a sign function. 
The corresponding part of the effective action
responsible for the generation of a CPTV mass spectrum according 
to this scenario reads:
\begin{eqnarray} 
S = \int d^4x {\bar \psi} i\partial_\mu \gamma^\mu 
\psi + \frac{im}{2\pi} \int d^3x dt dt' {\bar \psi}(t) \frac{1}{t - t'}
\psi (t') 
\label{actionnl} 
\end{eqnarray} 
Notice that in (\ref{actionnl}) Lorentz invariance is maintained 
(at least at tree level) due to the presence of the 
$\epsilon (p_0)$. However \emph{Locality} is relaxed, and hence CPT.
We stress 
once again that in this scenario the neutrino masses are of \emph{Dirac} type. 
An open issue, of course, 
of such scenaria for neutrino CPTV mass spectra 
is what singles out the neutrino sector from the rest
of the standard model so as to produce 
mass differences of the order of the LSND anomaly.

\begin{figure}[t]
\begin{center}
\includegraphics[width=.4\textwidth]{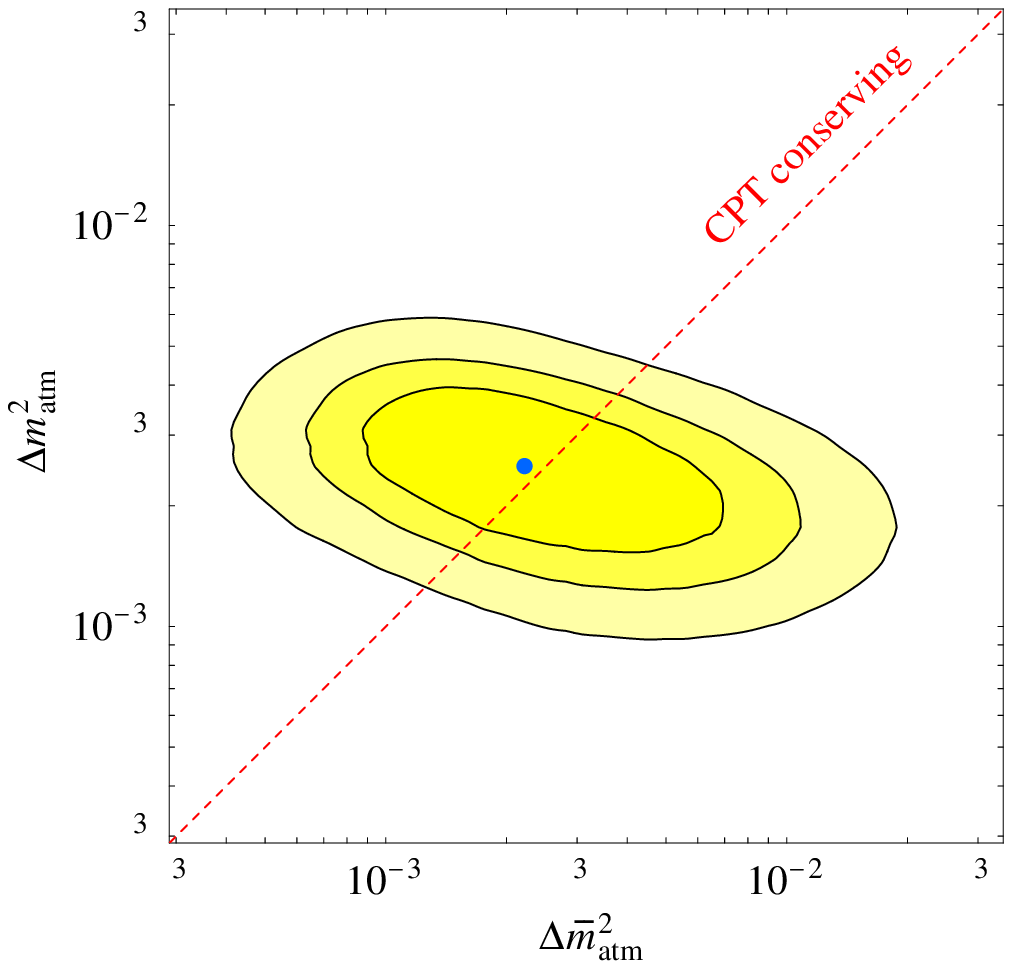}
\hfill \includegraphics[width=.4\textwidth]{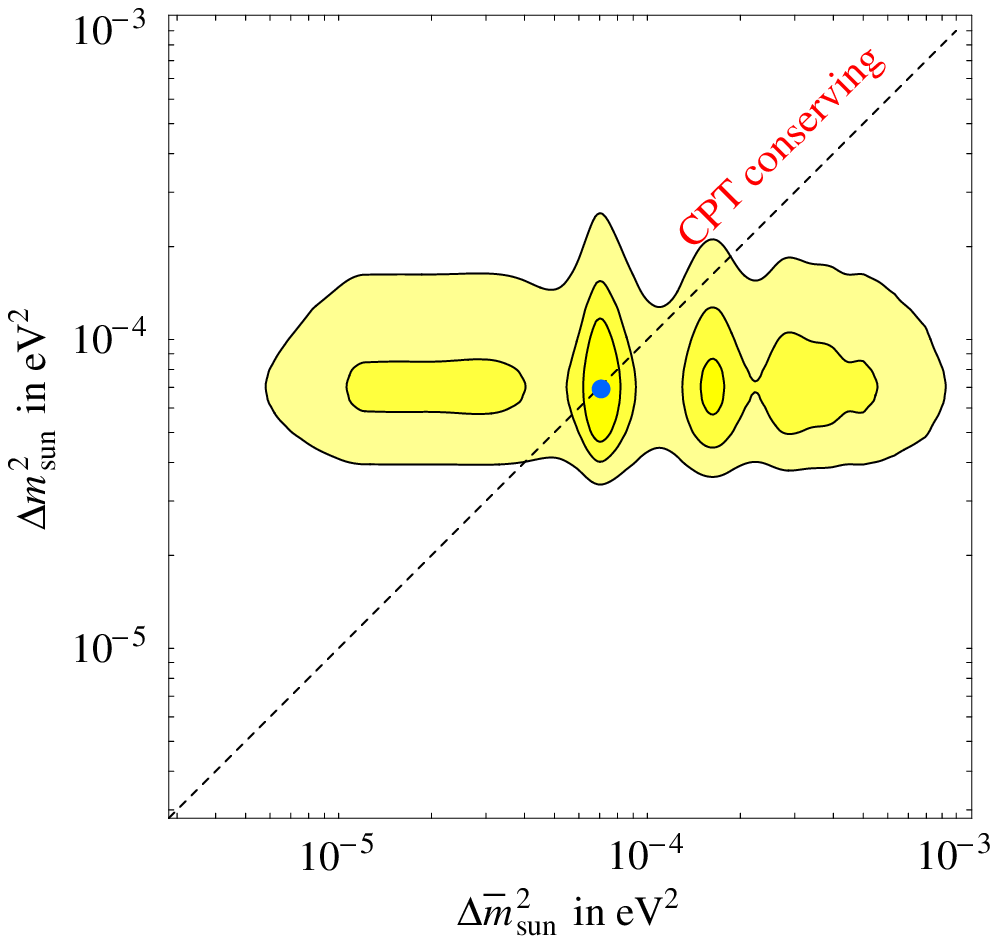}
\hfill \includegraphics[width=.4\textwidth]{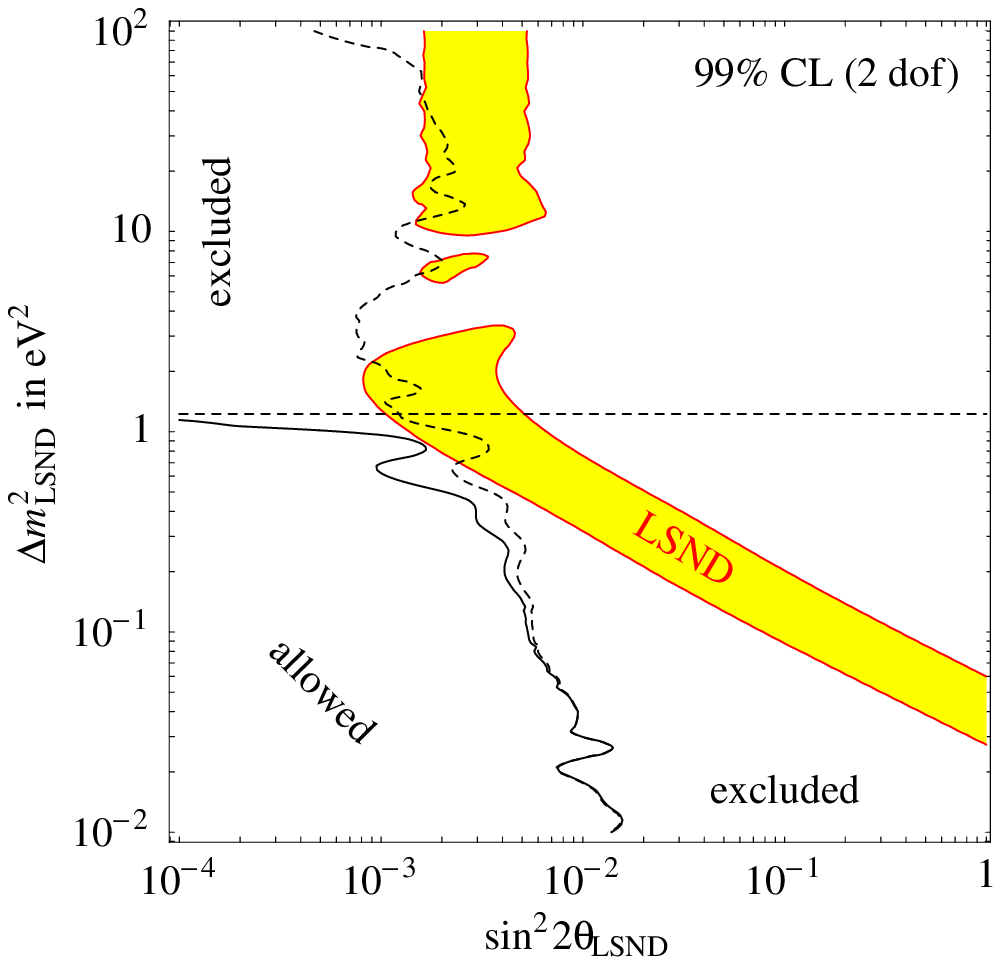}
\hfill \includegraphics[width=.4\textwidth]{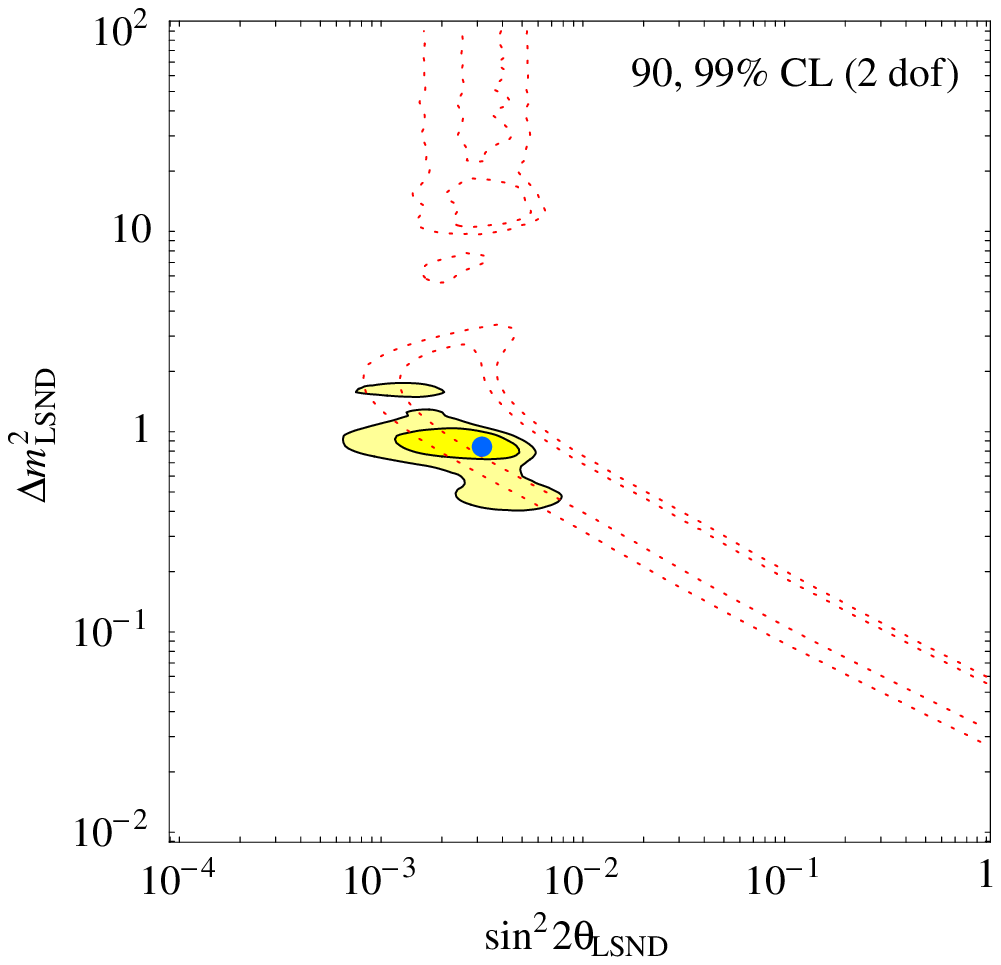}
\end{center}
\caption[]{The phenomenology of neutrino models 
involving CPT Violating mass spectra done in~\cite{strumia} seems to
be disfavored marginally when all the available current data 
are compiled. \underline{Upper figures: (a) Left}: 
Atmospheric $m_\nu - m_{\overline \nu}$ (68, 90, 99 \%, 2 d.o.f.).
\underline{(b) Right}: For solar and reactor data (68, 90, 99 \%, 2 d.o.f.). 
\underline{Lower Figures: (a) Left}: Upper half plane disfavored by WMAP. 
Dashed curved line: upper bound from all other $\nu$ experiments.
\underline{(b)Right}: Best fit, 
all data; 3 + 1 sterile $\nu$ solution disfavored by WMAP~\cite{wmap},
since $(\Delta m^2_{LSND})^{1/2} \simeq \sum m_\nu.$
\emph{Notation}: The oscillation probabilities 
are: $P_{\nu_e \to \nu_e} = 1 - S {\rm sin}^2 \theta _{es}~,~ 
P_{\nu_\mu \to \nu_\mu} = 1 - S {\rm sin}^2 \theta _{\mu s}~, 
P_{\nu_e \to \nu_\mu} = S {\rm sin}^2 \theta _{LSND}$, 
with ${\rm sin}^2\theta_{LSND} \simeq 
\frac{1}{4}{\rm sin}^2 2\theta_{es}{\rm sin}^2 2\theta_{\mu s}$. }
\label{strumiafig}
\end{figure}

The phenomenology of the model has been analyzed in~\cite{strumia}, 
where it was 
argued that a compilation of all available data 
from current neutrino experiments does 
not seem to favor CPT Violating scenaria for neutrino mass spectra.
The point is that if CPT is Violated in the neutrino sector, then such 
violations will imply signals in atmospheric and solar $\nu$ oscillations.
The analysis of \cite{strumia} combined recent results from 
KamLAND~\cite{kamland} with atmospheric data, as well as recent WMAP satellite 
data~\cite{wmap}. The results of such an analysis are 
given in fig. \ref{strumiafig}, from where it is evident that a 
CPTV $\nu$-mass scenario is excluded. 
 However, the conclusion is marginal, as can be seen from the 
figure,
and moreover it pertains only to CPT violation within conventional 
quantum mechanics, in the sense of a violation being realized through a 
neutrino-antineutrino mass difference. 

As we have discussed above, 
QG effects of the type appearing in the non-quantum mechanical 
evolution equations (\ref{evolution}) or 
(\ref{lindblad}),(\ref{liouvevoldm}) 
may be themselves responsible for neutrino oscillations,
in a way independent of mass terms~\cite{lisi,ben2}. 
Since such models 
violate CPT through the ill-definition of a scattering matrix,
as explained above, they may provide a natural explanation 
for the LSND anomaly, provided the latter 
is confirmed by future experiments. 
If this is the case, phenomenological analyses such as that in 
\cite{strumia} 
have to be redone by taking into account decoherence effects
in the dynamics of neutrinos.

Determining the 
order of such CPTV effects in the neutrino sector is not an easy task,
and is highly model dependent. However, if QG is responsible for the 
effects only through unitarity violation, 
and Lorentz invariance and locality are preserved, 
we expect that the order is (more or less) universal
among all particle species. 
As already mentioned, 
stringent bounds on such non quantum mechanical QG-induced 
decoherence 
parameters have been derived for the 
neutrino sector~\cite{lisi,ben2}, which are much smaller 
than the corresponding bounds in the Kaon sector~\cite{cplear}. 
This may guide us in theoretical modelling of these effects. 

\section{Conclusions and Outlook} 

In this brief review we considered 
some models of CPT violation and their
associated phenomenological constraints. From our discussion it becomes clear that in estimating 
the order of the effects, and hence the sensitivity of the various experiments,
it is important to know the details of the gravitational environment that may be 
responsible for such violations. There is no single figure of merit for
CPT violation, and hence detailed and systematic analyses have to be performed
with care before conclusions are reached. 

For instance, according to some non-critical 
stringy models of foam, the photons 
(and probably gluons) 
may exhibit modified dispersion relations, 
and hence LIV and CPTV, 
but such properties \emph{may not} characterize the rest of the 
particles in the standard model. This implies that future experiments
testing photon dispersion relations are important
in shedding light in the quantum nature of space time.

It may well be, of course, that CPT is broken, if at all, 
only cosmologically, 
in the sense discussed in the beginning of the article, in which case
any direct test via particle physics experiments seems  pointless due to 
the very small value of the associated effects. However, such effects may be 
tested indirectly by means of astrophysical observations, for instance 
experiments associated with a direct measurement 
of the acceleration of the Universe~\cite{snIa}, or  
measurements of the CMB anisotropies to a very high precision~\cite{wmap}. 
It is fascinating to link a possible confirmation of a cosmological 
\emph{constant} in our Universe 
with a breaking of CPT symmetry by a tiny but finite amount. 
Time will tell whether CPT symmetry is sacrosanct or follows the fate of 
so many other 
symmetries in nature, being broken by quantum space-time effects.
Fortunately such a question may be tackled by many experiments in the 
foreseeable 
future, especially those from the astrophysics side. 

Astrophysics has made enormous progress in improving 
the experimental sensitivities
over the past few years, which allows tiny numbers, such as 
the cosmological constant (if the latter is non zero), to be measured directly!
Thus, a fruitful co-operation with 
Particle Physics is to be expected for the exciting years to come.
In this review, we made an attempt 
to associate astrophysical/cosmological phenomena,
such as the Dark Energy content of the Universe, to fundamental 
concepts of Quantum Field Theory, such as CPT (non)invariance, and the 
structure of quantum space time. This may not be the only example,
where such a connection exists. Neutrinos, as we well as 
Supersymmetry searches, which we did not discuss here, 
constitute 
topics where new physics would definitely come into play, and where 
astrophysics has already given important results
and/or constraints~\cite{wmap,wmaprev}. 
One cannot also exclude 
the possibility of having pleasant surprises from 
antimatter factories, such as~\cite{cernhbar}, 
which may prove to be 
important for precision tests of CPT symmetry
in the not-so-distant future.

One therefore expects a plethora of experiments
from both Particle and Astrophysics side 
in the next decade at least, which 
hopefully will provide us with interesting physical results,
and thus enable us to make several steps forward  
in our quest for understanding the fundamental interactions in Nature. 
I would like to close this discussion with a remark made by Okun in a 
lecture on CPT symmetry~\cite{okun}: if CPT is broken,
he said, then the whole structure on which we built the 
current form of quantum field theory, and on which our 
phenomenology is based, may cease to exist. How can we proceed then,
so as to make sure that we detect and interpret such a violation correctly?
As we have 
discussed in this review, this may indeed turn out to be true,
but it may not be so drastic as one thinks. The dynamics of 
open systems, for instance, familiar from other fields of physics, such as condensed matter, may be the way forward. Time will tell... 

\section*{Acknowledgments} 

The author thanks H. Klapdor-Kleingrothaus for the invitation to 
the Conference \emph{Beyond the Desert 2003}. The work of N.E.M. 
is partially supported by a Visiting Professorship at the 
Department of Theoretical
Physics of the University of Valencia (Spain), and by the European Union
(contract HPRN-CT-2000-00152).

%

\end{document}